\documentclass[aps,prd,twocolumn,groupedaddress,nofootinbib,showpacs,eqsecnum,useAms]{revtex4}
\usepackage{epsfig}
\usepackage{euscript,graphicx}
\usepackage{amsthm,dsfont,amsfonts,amsmath,amssymb}
\usepackage{euscript,color,fontenc,textcomp,relsize}
\usepackage{bm,url,float,cleveref}
\usepackage[caption=false]{subfig}
\usepackage{natbib}

\allowdisplaybreaks
\def\no{\nonumber \\}
\def\noo{\no&&\,}
\def\bk {\\ & \quad}
\def\bes{\begin{subequations}}
\def\ens{\end{subequations}}
\newcommand{\bea}{\begin{eqnarray}}
\newcommand{\eea}{\end{eqnarray}}

\newcommand{\vek}[1]{\boldsymbol{#1}}

\begin{document}

\date{\today}


\title{ Gravitational waves from compact binaries in post-Newtonian accurate hyperbolic orbits}

\author{ Gihyuk Cho\,$^{1}$\footnote{whrlsos@snu.ac.kr},
 Achamveedu Gopakumar\,$^{2}$, 
 Maria Haney\,$^3$, Hyung Mok Lee\,$^4$}

\affiliation{$^{1}$
Department of Physics and Astronomy,
Seoul National University,
Seoul 151-742, Korea \\
$^2$Department of Astronomy and Astrophysics,
 Tata Institute of Fundamental Research, Mumbai 400005, India
 \\
 $^3$Physik-Institut, 
Universit\"{a}t Z\"{u}rich,
Winterthurerstrasse 190,
 8057 Z\"{u}rich\\
$^4$Korea Astronomy and Space Science Institute, Dajeon, Korea
}

\begin{abstract}

We derive from first principles third post-Newtonian (3PN) accurate
Keplerian-type parametric  solution to describe PN-accurate dynamics of 
non-spinning compact binaries in hyperbolic orbits.
Orbital elements and functions of the parametric solution are obtained 
in terms of the conserved orbital energy and angular momentum in 
both Arnowitt-Deser-Misner type and modified harmonic coordinates. 
Elegant checks are provided that include a modified analytic continuation prescription to 
obtain our independent hyperbolic parametric solution from its eccentric version.
A prescription to 
model gravitational wave polarization states 
for hyperbolic compact binaries experiencing  3.5PN-accurate orbital motion is presented
that employs our 3PN-accurate parametric solution.

\end{abstract}


\maketitle

\section{Introduction}

Interesting astrophysical scenarios involving strong gravitational
fields usually require accurate and efficient ways of describing  
orbital dynamics of compact binaries.
These scenarios include gravitational wave (GW) events,  
observed by the advanced LIGO -Virgo interferometers \cite{aLIGO}, 
labeled GW150914, 
GW151226, GW170104, GW170608, GW170814, and GW170817 \cite{GW_1_d, GW_2_d,GW_3_d, GW_6_d,GW_4_d, GW_5_d}.
The first five are  associated with the coalescence of 
black hole (BH) binaries while GW170817 involved a merging neutron star binary.
The other strong field scenarios involving compact binaries include 
radio observations of relativistic binary pulsars like PSR~1913+16 and PSR~J0737-3039 \cite{JT_1993,J0737}
and optical observations of Blazar OJ287, powered by
a massive BH binary central engine \cite{MV2016}.

Orbital dynamics of compact binaries spiraling in due to the emission of GWs  can be accurately
 described by the PN approximation to general relativity \cite{LR_LB}.
In this approximation, orbital dynamics of 
 non-spinning compact binaries  is provided 
as corrections to Newtonian equations of motion
in powers of $ ( v/c)^2 \sim G\, m/(c^2\,r)$, where $v, m,$ and $r$ are
the velocity, total mass and relative separation of the binary.
At present, conservative orbital dynamics of compact binaries have been computed to the fourth PN order
which provides $(v/c)^8$ accurate 
general relativity based corrections to Newtonian description
 (see for example Refs.~\cite{Porto17,DJ17,Foffa16,BBBFM16,DJS2016} and their many references 
for a glimpse of this herculean effort from various approaches).
Interestingly, it is possible to obtain a Keplerian-type parametric 
solution to PN-accurate orbital dynamics of compact binaries
in non-circular orbits.
This was demonstrated by Damour and Deruelle 
for 1PN-accurate compact binary orbital dynamics,
relevant for both eccentric and hyperbolic orbits \cite{DD1985}.
They introduced three eccentricities so that the 
parametrization looks `Keplerian' even at 1PN order.
These computations were extended to 2PN and 3PN orders 
by Sch\"afer and his collaborators which led to the 
 generalized quasi-Keplerian parametric 
solution for compact binaries in precessing eccentric orbits \cite{DS1988,SW93,MGS}.
This solution plays an important role 
in the on-going efforts to model GWs from 
merging BH binaries in eccentric orbits  \cite{Hinder_2010,eIMR}.
This is due to the use of 
certain GW phasing formalism, 
developed in Refs.~\cite{DGI,KG06}, for describing the 
inspiral part of eccentric binary coalescence.
This formalism employs 
Keplerian-type 
parametric solution to model  orbital and periastron precession timescale
variations present in the two GW polarization states $h_+(t) $ and $h_\times(t)$.
These features are crucial to obtain 
 $h_+(t) $ and $h_\times(t)$
from compact binaries inspiraling along PN-accurate eccentric orbits in an accurate 
and efficient manner \cite{THG16}.
Additionally,  high precision radio
observations of binary pulsars employ an accurate relativistic `timing
 formula' \cite{DD86,DT92}  which requires
1PN-accurate Keplerian type parametric solution 
for compact binaries moving in precessing eccentric orbits 
\cite{DD1985}.
This timing formula  is crucial to 
test both the predictions of general relativity and the 
viability of alternate theories of gravity in strong field situations present in our galaxy
\cite{IS03_LR}.

In this paper, we  derive from first principles  parametric solution to 3PN accurate 
conservative orbital dynamics  of  compact binaries 
moving in hyperbolic orbits.
This parametric solution is given both in 
Arnowitt, Deser, and Misner (ADM)- type and modified harmonic (MH) coordinates. The reason that we adpat both gauges is that ADM is useful for comparing with numerical data from  numerical relativity (NR)  simulations which make use of ADM formalism, and MH is proper for constructing GW waveforms.
The associated orbital elements and functions are 
provided as PN-accurate functions of the conserved 
orbital energy, angular momentum and the symmetric mass ratio.
The correctness of our solutions is verified by comparing 3PN-accurate 
expressions for the radial and angular velocities 
arising from our solutions with their counterparts, computed directly 
from the orbital dynamics.
Additionally, we develop 
a modified analytic continuation prescription to 
obtain our  3PN-accurate Keplerian 
type parametric solution for hyperbolic orbits from its 
 eccentric versions, available in Ref.~\cite{MGS}.
 This is a desirable feature as we are essentially providing an  
additional test on the correctness of lengthy expressions present in  Ref.~\cite{MGS}
which are, as noted earlier, required to construct templates for eccentric inspirals. 
We also obtain temporally evolving GW polarization states for  compact
binaries in 3.5PN-accurate hyperbolic orbits. 
This is achieved by 
allowing orbital elements and functions of our  3PN accurate Keplerian 
type parametric solution to vary due to 1PN-accurate radiation reaction effects,
relevant for hyperbolic orbits \cite{KG06,LGGJ}.
Our efforts are motivated by the observation that 
compact binaries in unbound orbits are 
plausible GW sources for both the ground and space based GW observatories.
It turned out that such  rare events are expected to occur in 
globular clusters and galactic nuclear clusters or plausibly in 
dense clusters of primordial black holes \cite{KGM06,BN17}.
Moreover,  hyperbolic encounters may 
 create bound binaries having very high eccentricities \cite{H72,WW79}.
 It was argued that 
plausible  detection rates for such eccentric binaries may become 
comparable to that for  isolated compact binary coalescences  \cite{OKL09}.
 Interestingly, such hyperbolic GW events involving neutron stars may even 
be accompanied by electro-magnetic flares  \cite{DT13}.
The present effort should provide accurate gravitational waveforms for hyperbolic 
passages that can be adapted in to the 
LSC Algorithm Library Suite of the LIGO Scientific Collaboration.

  The paper is organized in the following way. In Sec.~\ref{sec:I} we first present 
 the derivation of 1PN-accurate Keplerian-type parametric solution for 
 hyperbolic orbits from first principles.
 This is followed by detailing our approach to extend it to 3PN order 
 and ways to check the correctness of our solutions in two different gauges.
 We present an accurate and efficient way  to obtain temporally evolving GW polarization states 
for non-spinning compact binaries moving in 3.5PN accurate hyperbolic orbits in Sec.~\ref{sec:II}.
A brief summary and  possible extensions  are listed in Sec.~\ref{sec:conclusions}.

\section{Derivation of Keplerian type parametric solution for  compact binaries in hyperbolic orbits}
\label{sec:I}
We first provide a detailed derivation of a 1PN-accurate Keplerian type parametric solution for compact binaries in hyperbolic 
orbits. This procedure explicitly demonstrates why one requires three 
eccentricity parameters to obtain the desired solution, whhich was previously computed by employing 
certain analytic continuation arguments in Ref.~\cite{DD1985}.
The 3PN extension of Sec.~\ref{secIIa} is detailed  in Sec.~\ref{secIIb}.

\subsection{1PN-accurate quasi-Keplerian parametrization for hyperbolic orbits}
\label{secIIa}

 We begin by displaying  1PN-accurate expressions for 
 the radial and angular motion 
 \begin{subequations}
 \label{radial equation}
\begin{align}
\biggl( \frac{d r}{dt} \biggr )^2 & = a_0 + a_1\,s + a_2\,s^2 + a_3\,s^3 \,,\\ 
\frac{d\phi}{dt} &= d_0 \, s^2 + d_1\, s^3 \,, 
\end{align}
\end{subequations}
where both radial and temporal variables are scaled by $G\,m$ \cite{DD1985}.
This allows us to introduce a variable 
 $s = 1/r$, where $ r= | \vek R|/(G\,m)$ and $\vek R$ is the relative separation vector 
such that $\vek R = R\, ( \cos \phi, \sin \phi, 0)$.
The constant coefficients, $a_0,a_1,a_2,a_3, d_0$ and $d_1$
are given in terms of certain conserved orbital energy ($\tilde E$) and angular momentum ($\tilde J$) as
\bes
\label{Eq.2.2}
\setlength{\jot}{10pt}
\begin{align}
a_0 =&\, \frac{2\tilde E}{\mu}+\frac{1}{c^2}\frac{(2\tilde E)^2}{\mu^2}\frac{-3+9\eta}{4}
\,,\\
a_1 =&\, 2-\frac{1}{c^2}\frac{2\tilde E}{\mu}(6-7\eta)
\,,\\
a_2 =&\, -\frac{\tilde J^2}{G^2m^2\mu^2}-\frac{1}{c^2}\Big[\frac{ 2\tilde E \tilde J^2}{G^2m^2\mu^3} (3 \eta-1)\no &+(-10 + 5 \eta)\Big]
\,,\\
a_3 =&\, \frac{1}{c^2}\frac{\tilde J^2}{G^2m^2\mu^2} (8-3 \eta) 
\,,\\
d_0 =&\,\frac{\tilde J}{Gm\mu}+\frac{1}{c^2}\frac{2\tilde E\tilde J}{G^2m^2\mu^3}\frac{-1+3\eta }{2}
\,,\\
d_1=&\,\frac{1}{c^2}\frac{\tilde J}{Gm\mu}(-4+2\eta)
\,,
\end{align}
\ens
where  $\mu $ and $\eta$ denote the  usual reduced mass and symmetric mass ratio.
It should be obvious that these coefficients take simpler forms in terms of certain reduced 
 orbital energy and angular momentum variables,  defined as   $E = \tilde E/\mu, h = \tilde J/(G\,m\, \mu)$. 
Additionally,  we are considering unbound hyperbolic orbits, therefore  $E>0$.

Influenced by Ref.~\cite{DD1985}, 
 we tackle the radial motion by introducing a certain {\it conchoidal} transformation 
\begin{align}
r = \bar r + \frac{a_3}{2\,a_2'}
\end{align}
where $a_2'=-h^2$ so that $a_3/ 2\,a_2' \sim O(\frac{1}{c^2})$ and $\lim_{c^{-1}\rightarrow 0} {a_2}$ gives $a_2'$.
It is fairly straightforward to recast the above radial equation in terms of $\bar r$ as 
\begin{align}
\biggl (\frac{d\bar r}{dt}\biggr )^2 
&= a_0+\frac{a_1}{\bar r}+\frac{a_2}{\bar r^2}-\frac{a_1a_3}{2\,a'_2\,\bar{r}^2}+O(\frac{1}{c^4})\,, 
\nonumber \\
& =a_0 +\frac{a_1}{\bar r}+\frac{\bar a_2}{\bar r^2}+O(\frac{1}{c^4})\,,
\end{align}
where $\bar a_2 = a_2-\frac{a_1a_3}{2a'_2}$, while consistently neglecting terms ${\cal O}(\frac{1}{c^4})$.
To solve the above equation, we introduce an angular parameter 
$u$ such that 
\begin{align}
\label{Eq.2.5}
\frac{du}{dt} = \frac{1}{ a_4\, \bar r} >0 \quad; \quad  a_4>0.
\end{align}
This leads to 
\begin{align}
\biggl ( \frac{d\bar r}{du} \biggr)^2 = a_4^2 \left ( a_0\bar r^2 +a_1 \bar r+\bar a_2 \right )\,.
\end{align}
Clearly, we require  $(a_0\,\bar r^2 +\,a_1\, \bar r+ \bar a_2) >0$,  and this 
allows us to write 
\begin{align}
 \pm a_4\, du
 & =\frac{d\bar r}{\sqrt{a_0\bar r^2 +a_1 \bar r+\bar a_2}}
\,\\  \nonumber
& =\frac{d\bar r}{\sqrt{\bar a_2-\frac{a_1^2}{4a_0}+a_0(\bar r+\frac{a_1}{2a_0})^2 }}\,.
\end{align}
For hyperbolic motion, we observe that $\left ( \bar a_2- a_1^2/(4\,a_0 ) \right )$ is indeed negative.
Therefore, we re-write this equation as 
\begin{align}
\pm a_4 \, \left( \sqrt{-\bar a_2+\frac{a_1^2}{4a_0}} \right) \,du = \frac{d\bar r}{\sqrt{-1+\frac{4a_0^2}{a_1^2-4\bar a_2 a_0}(\bar r+\frac{a_1}{2a_0})^2 }}\,.
\end{align}
We now introduce $u'$ such that  $\cosh u'= \sqrt\frac{4a_0^2}{a_1^2-4\bar a_2 a_0}(\bar r +\frac{a_1}{2a_0}) $ 
which allows us to simplify  
the above equation as 
\begin{align}
\pm a_4\,\sqrt{ a_0}du = du'\,.
\end{align}
We let 
 $a_4=\frac{1}{\sqrt{a_0}}$ to ensure that  $\cosh u' = \cosh (\pm u)=\cosh u$.
 This leads to the following equations for $\bar r$ as well as $r$ 
\bes
\setlength{\jot}{10pt}
\begin{align}
\bar r& = -\frac{a_1}{2a_0}+\sqrt\frac{a_1^2-4\bar a_2a_0}{4a_0^2} \cosh u\,, \\
r &=  -\frac{a_1}{2a_0}+\frac{a_3}{2a'_2}+\sqrt\frac{a_1^2-4\bar a_2 a_0}{a_0^2} \cosh u\,,
\nonumber  \\
&= \left ( \frac{a_1}{2a_0} -\frac{a_3}{2a'_2} \right )\, \biggl [
 -1 +  \left ( \frac{a_1}{2a_0} -\frac{a_3}{2a'_2} \right )^{-1}\,
\nonumber  \bk 
 \times \,  
 \sqrt\frac{a_1^2-4\bar a_2 a_0}{4a_0^2} \cosh u \biggr ]\,.
\end{align}
\ens
We now identify $\left ( \frac{a_1}{2a_0} -\frac{a_3}{2a'_2} \right )$ with $a_r$
and the coefficient of $\cosh u$ with $e_r$.
The 1PN-accurate expression for $e_r$ is therefore given by  
\begin{align}
e_r& = \left (\frac{a_1}{2a_0}-\frac{a_3}{2a'_2} \right )^{-1}\sqrt{\frac{a_1^2-4\bar a_2 a_0}{4\,a_0^2}}\,
\nonumber \\
&=(1+\frac{a_0\,a_3}{a_1\,a'_2} )\sqrt{1-\frac{4\bar a_2a_0}{a_1^2}}+O(\frac{1}{c^4})\,.
\end{align}

 Invoking Eqs.~(\ref{Eq.2.2}), the parametric equation for $r$ may be summarized as 
\bes
\setlength{\jot}{10pt}
\label{r_1PN}
\begin{align}
r =& \,a_r \,\left ( e_r\, \cosh u - 1 \right )\,,\\
a_r =\,&   \frac{1}{2E}+\frac{1}{c^2}(\frac{7}{4}-\frac{\eta}{4})
\,,\\
e_r^2 =& \,1+2Eh^2\no
&+\frac{1}{c^2}(2E)\biggl [ -44+19\eta+2Eh^2\, \left ( \frac{1}{2}-\frac{\eta}{2} \right )\biggr ]
\,,
\end{align} 
\ens
We have verified that the expression for $a_r$ is identical to Eq.~(7.4) in Ref.~\cite{DD1985}, obtained 
by invoking the argument of analytic continuation. 

 To obtain the 1PN accurate Kepler equation for hyperbolic orbits, we turn to Eq.~(\ref{Eq.2.5}) for the angular variable $u$ 
and integrate it. This leads to 
\begin{align}
\sqrt{a_0}(t-t_0) & =\int \bar r du \,, \nonumber \\ 
& =\int du \biggl (-\frac{a_1}{2a_0}+\sqrt\frac{a_1^2-4\bar a_2 a_0}{4a_0^2} \cosh u \biggr) \,, \\ \nonumber
& = \biggl ( -\frac{a_1}{2a_0}u+\sqrt\frac{a_1^2-4\bar a_2 a_0}{4a_0^2} \sinh u \biggr ) \,.
\end{align}

 It is straightforward to re-write the above equation in its more familiar form 
\bes 
\label{ke_1PN_a}
\begin{align}
n(t-t_0) &= e_t\sinh u-u \,,\,\,\, \mbox {where} \\
n &= \frac{2a_0^{\frac{3}{2}}}{a_1}
\,,\\
e_t &= \sqrt{1-\frac{4\bar a_2a_0}{a_1^2}}
\,.
\end{align}
\ens
Using Eqs.~(\ref{Eq.2.2}), we can express the orbital elements  $n$ and $e_t^2$
in terms of the conserved $E,h$ and $\eta$ as
\bes
\label{ke_1PN_b}
\begin{align}
n &=(2E)^\frac{3}{2}+\frac{1}{c^2}(2E)^\frac{5}{2}\frac{15-\eta}{8}
\,,\\
e_t^2 &=1+2Eh^2\\\notag
&+\frac{1}{c^2}(2E)\big[-18+8\eta+2Eh^2(\frac{17}{4}-\eta\frac{7}{4})\big]
\,.
\end{align}
\ens
The above expressions are also identical to those given in Ref.~\citep{DD1985}.

 We are now in a position to tackle the angular motion.
 Influenced by Ref.~\citep{DD1985},
we employ another conchoidal transformation
\begin{align}
\tilde r
& = r - \frac{d_1}{2d_0}\,.
\end{align}
Using our expression for $r$, we obtain 
\begin{align}
\tilde r 
& =\tilde a(\tilde e\cosh u-1)\,,
\end{align}
where $\tilde a = a_r -\frac{d_1}{2d_0}$ and $\tilde e = \frac{a_r\,e_r}{\tilde a}$.
In terms of $\tilde r$, 
the 1PN-accurate equation for the angular motion, given in Eqs.~(\ref{radial equation}), takes the simpler Newtonian form:
\begin{align}
\frac{d\phi}{dt} = \frac{d_0}{\tilde r^2}\,.
\end{align}
With the help of our 1PN-accurate Kepler Equation, this leads to 
\begin{align}
\frac{d\phi}{du} = \frac{d_0}{n\,\tilde a^2} \frac{( e_t\cosh u-1)}{(\tilde e\cosh u-1)^2}\,.
\end{align}
We introduce $e_\phi = 2\,\tilde e- e_t$ which allows us to simplify 
$ ( e_t\cosh u-1 )/ ( (\tilde e\cosh u-1)^2 )$ as 
$1/( e_\phi\cosh u-1 )$, modulo the neglected  ${\cal O}({1/c^4})$ terms.
 Integrating the resulting expression for $ d \phi/ du = d_0/ ( n\, \tilde a^2\, ( e_\phi\cosh u-1 ))$ gives us 
\begin{align}
\phi - \phi_0 
&=  \frac{d_0}{n\tilde a^2}\int \frac{du}{e_\phi\cosh u-1} \nonumber \\
& = \frac{d_0}{n\tilde a^2}\int \frac{du}{(e_\phi-1)\cosh^2 \frac{u}{2}+(e_\phi+1)\sinh^2 \frac{u}{2}}
\nonumber \\
& =\frac{d_0}{n\, \left ( \sqrt{e_\phi^2-1} \right )\,\tilde a^2}
\,\int \frac{du\sqrt\frac{e_\phi+1}{e_\phi-1}\frac{1}{\cosh^2 \frac{u}{2}}}{1+\frac{e_\phi+1}{e_\phi-1}\tanh^2 \frac{u}{2}}
\nonumber \\
& =\frac{d_0}{n\, \tilde a^2\, \sqrt{e_\phi^2-1}}\, 2\arctan(\sqrt\frac{e_\phi+1}{e_\phi-1}\tanh \frac{u}{2})\,.
\end{align}
We now introduce $K$ such that 
\bes
\label{phi_1PNa}
\begin{align}
\phi - \phi_0  &=  K \times 2\, \arctan(\sqrt\frac{e_\phi+1}{e_\phi-1}\tanh \frac{u}{2})\,,\,\,\, \mbox{where}
\\ 
K &= \frac{d_0}{n\, \tilde a^2\, \sqrt{e_\phi^2-1}}\,,
\end{align}
\ens
It is straightforward to express the orbital elements $K$ and $e_\phi$ in terms of the conserved 
quantities like $E$ and $h$, and we have 
\bes
\label{phi_1PNb}
\begin{align}
K &= 1+\frac{1}{c^2}\frac{34-15\eta+(2Eh^2)(-8+3\eta)}{2h^2}
   \,,\\
e_\phi^2 &=1+2Eh^2\\\notag
&+\frac{1}{c^2}(2E)\big[-34+15\eta+2Eh^2(\frac{-47}{4}+\frac{21}{4}\eta)\big]
\,.   
\end{align}
\ens

 We have verified that Eqs.~(\ref{r_1PN}), (\ref{ke_1PN_a}, (\ref{ke_1PN_b}), (\ref{phi_1PNa}) 
and  (\ref{phi_1PNb}) 
 indeed are identical to their 
counterparts in Ref.~\cite{DD1985} that have been obtained by invoking the arguments of  analytic continuation. 
These arguments establish that the parametric elliptic solution (i.e., in the case of $E<0$) is well defined and analytic in $E$ and $u$ even in the domain 
where $E >0$ and 
$u \rightarrow i\, u$ is purely imaginary.
The solution for the 1PN-accurate hyperbolic motion obtained in Ref.~\cite{DD1985} through analytic continuation to $E>0$ and purely imaginary $u$ 
has been found to be thhe same as  Eqs.~(\ref{r_1PN}), (\ref{ke_1PN_a}, (\ref{ke_1PN_b}), (\ref{phi_1PNa}) 
and  (\ref{phi_1PNb}). 
Note that the three distinct eccentricity parameters $e_t$, $e_\phi$, $e_r$ ensure that the 1PN-accurate parametric solution looks 
quasi-Keplerian. The presence of $K$ can modify the trajectories of PN-accurate hyperbolic orbits 
with respect to their Newtonian counterparts, as we will demonstrate later.
In the next subsection, we extend these calculations to  3PN order. 

\subsection{ 3PN-accurate hyperbolic generalized quasi-Keplerian 
parametrization for compact binaries }
\label{secIIb}

The plan is to derive from first principles a 3PN-accurate Keplerian-type parametric solution 
for compact binaries in hyperbolic orbits. 
We are attempting the 3PN extension of Sec.~\ref{secIIa}, as it is not straightforward to obtain a hyperbolic 
version of the 3PN-accurate generalized quasi-Keplerian 
parametrization for compact binaries in eccentric orbits, detailed in Ref.~\cite{MGS}, 
simply by invoking the analytic continuation arguments of  Ref.~\cite{DD1985}.
The main difficulty with analytic continuation is due to the structure of 3PN-accurate (eccentric) Kepler Equation, given by Eq.~(19b) in Ref.~\cite{MGS},
which reads
\begin{align}
l \equiv& n' \left ( t - t_0 \right ) =
u' -e_t\,\sin u' +
\left ( \frac{g_{4t'}}{c^4} + \frac{g_{6t'}}{c^6} \right )\,(v' -u')
 \no
 &
  +
  \left ( \frac{f_{4t'}}{c^4} + \frac{f_{6t'}}{c^6} \right )\,\sin v'
  + \frac{i_{6t'}}{c^6} \, \sin 2\, v'
  + \frac{h_{6t'}}{c^6} \, \sin 3\, v' \,,
\end{align}
where $n',u'$ and $v'$ stand for the mean motion, eccentric and true anomalies of an eccentric orbit, and where
the orbital functions $ g_{4t'},\,g_{6t'},\, f_{4t'},\,f_{6t'},\, i_{6t'},\, h_{6t'}$ are PN-accurate functions of the conserved 
energy, angular momentum and the symmetric mass ratio $\eta$.
It is customary to employ the following exact expression for $(v'-u')$, derived in Ref.~\cite{KG06}:
\begin{align}
\label{vmu_eq}
v' - u' & = 2 \tan^{-1}
\left(
\frac{ \beta'_{\phi} \sin u' }{ 1 - \beta'_{\phi} \cos u' }
\right)
\,,
\end{align}
where $\beta'_{\phi} = ( 1 - \sqrt{ 1 - e_\phi^2 } ) / e_\phi$.
A close inspection reveals that it is certainly problematic to apply the usual 
analytic continuation arguments of \cite{DD1985}, namely, to let $ u' \rightarrow \imath v$ and 
allow $\sqrt{-E} \rightarrow \imath\,\sqrt{E}$ 
to obtain the hyperbolic version of an exact expression for $v'- u'$.
Additionally, the presence of $(-2\,E\, h^2)^{\frac{1}{2}}$ and its multiples in the explicit 
expressions for $n', g_{4t'}$ and $g_{6t'}$, as given by Eqs.~(20) of Ref.~\cite{MGS}, introduces further complications while trying to 
achieve hyperbolic versions of these expressions.

  These considerations prompted us to obtain hyperbolic versions of Eqs.~(19), (20) and (21)
of Ref.~\cite{MGS} (describing the radial and angular motion in an eccentric binary, as well as the Kepler equation) with the help of ab-initio computations.
It turned out that these detailed computations enabled us to 
devise a modified version of the standard analytic continuation arguments, 
 in order to extract hyperbolic counterparts of the expressions in Ref.~\cite{MGS}.
This allowed us to check the correctness of our computations 
and to confirm the validity of Ref.~\cite{MGS}.
An additional way of checking the lengthy expressions in 
 Ref.~\cite{MGS} is highly desirable, as this work is usually invoked 
for the GW phasing of compact binaries inspiraling along eccentric orbits.

 We begin by tackling the hyperbolic radial part of the 3PN-accurate Keplerian-type parametric solution.
 The input for our calculation is the following 3PN-accurate expression for $\dot r^2$,
symbolically written as  
 \begin{align}
\label{Eq_dtr_s3}
\dot r ^2 \equiv & \frac{1}{s^4}
{\left({\frac {{ ds}}{{ dt}}}\right)}^{2}\\\notag=& a_{{0}}+a_{{1}}s+
a_{{2}}{s}^{2}+a_{{3}}{s}^{3}+a_{{4}}{s}^{4}+a_{{5}}{s}^{5}+a_{{6}}{s}
^{6}+a_{{7}}{s}^{7} \,,
\end{align}
where explicit functional forms of the coefficients $a_i$ are given by Eqs.~(A1) and (A3) of Ref.~\cite{MGS}, for the ADM-type 
and modified harmonic gauges, respectively. 
We observe that in the Newtonian limit the right hand side of Eq.~(\ref{Eq_dtr_s3}) is 
a second order polynomial in $s$ and therefore admits two roots.
It is straightforward to obtain 3PN-accurate versions of these two real-valued roots, even in the case of hyperbolic orbits.
Subsequently, we factorize the 3PN-accurate expression for $\dot r^2$ with the help of 
the two roots $s_+$ and $ s_-$. This leads to
\begin{align}
(t-t_0)& = \int\frac{ds\,\, \left( b_0+b_1s+b_2s^2+b_3s^3+b_4s^4+b_5s^5 \right )}{s^2\sqrt{(s_+-s)(s-s_-)}}\,,
\end{align}
where we used the parametric equation $r =  a_r(e_r\cosh u -1)$. The explicit functional forms for the coefficients $b_i$ may be  found in Eqs.~(A2) and (A4) of Ref.~\cite{MGS}.  Note the factorization of the denominator and 
how it differs from Eq.~(9) of Ref.~\cite{MGS}. This is because for hyperbolic orbits we find the roots 
$s_+ >0$ and $s_-<0$, which allows us to introduce $r =  a_r(e_r\cosh u -1)$.
The above integral leads to
\begin{align}
\label{tt0_ciprime}
t-t_0  =& \, c'_0\,(e_r\sinh u -u)+c'_1\, u+\frac{c'_2\, \nu'}{\sqrt{e_r^2-1}}\\\notag
& +\frac{c'_3}{(e_r^2-1)^{3/2}} \biggl( \nu'+e_r\sin\nu' \biggr) \\\notag
&+ \frac{c'_4}{(e_r^2-1)^{5/2}} \biggl (\frac{e_r^2+2}{2}\nu'+2e_r\sin\nu'+\frac{e_r^2}{4}\sin2\nu' \biggr )\\\notag
 &+\frac{c'_5}{(e_r^2-1)^{7/2}}\biggl[ (1+\frac{3e_r^2}{2})\nu'+(3e_r+\frac{3}{4}e_r^3)\sin\nu'\\\notag
&\quad +\frac{3e_r^2}{4}\sin2\nu'+\frac{e_r^3}{12}\sin3\nu' \biggr ]\,,
\end{align} 
where $c'_i = b_i/( a_r^{i-1}\, \sqrt{-s_+\, s_-})$
and $ \nu'= 2\frac{1}{\sqrt{e_r^2-1}}\arctan(\sqrt{\frac{e_r+1}{e_r-1}}\tanh\frac{u}{2})$.
The above equation can be re-written as
\begin{align}
\label{temp_t}
t-t_0 &=c_0\sinh u-c_1\,u +c_2\nu'+c_3\sin\nu'\\\notag
&+c_4\sin2\nu'+c_5\sin3\nu' \,, 
\end{align}
with explicit relations between the coefficients $c_i$s and $c_i'$s  given in Appendix~\ref{AppC}.

It is straightforward to deduce that the coefficient $c_3$ of $\sin \nu'$ in Eq.~(\ref{temp_t}) begins at 1PN order.
Therefore, the above result deviates from our 1PN-accurate Keplerian-type parametric solution, derived 
in the previous section. It turns out that a suitable change of the $\nu'$ variable can remedy this undesirable 
feature, which will be addressed later.

 We turn our attention to the angular motion. The relevant ingredient of the calculation is $d \phi/d s= \dot \phi/\dot s$, which may be 
 symbolically written as 
\begin{align}
\frac{d\phi}{ds} = -\frac{d_0+d_1s+d_2s^2+d_3s^3+d_4s^4+d_5s^5}{\sqrt{(s_+-s)(s-s_-)}}\,,
\end{align}
where the coefficients $d_i$  are listed in Eqs.~(A2) and (A4) in Ref.~\cite{MGS} (there denoted as $B_i$), for ADM-type and 
modified harmonic gauges, respectively.
This leads to 
\begin{align}
\label{phiphi0_eiprime}
\phi-\phi_0 = \int  du \biggl ( \frac{e'_0}{r}+\frac{e'_1}{r^2}+\frac{e'_2}{r^3}+\frac{e'_3}{r^4}
+\frac{e'_4}{r^5}+\frac{e'_5}{r^6} \biggr )\,,
\end{align}
where $e'_i = d_i/( a_r^{i+1}\,\sqrt{- s_+s_-})$.
The above expression can be integrated to obtain 
\begin{align}
\label{Eq.2.29}
 \phi-\phi_0&= e_0\nu'+e_1\sin\nu'+e_2\sin2\nu'+e_3\sin3\nu'\\\notag&+e_4\sin4\nu'+e_5\sin5\nu' \,.
\end{align}
As expected, the coefficients $e_i$ are certain PN-accurate expressions and are given as functions of $e_i'$ in Appendix~\ref{AppC}. We observe that the coefficient of 
$\sin \nu'$ in Eq.~(\ref{Eq.2.29}), namely $e_1$, begins at 1PN order. Therefore, the above functional form for the angular motion $\phi- \phi_0$ 
also  deviates from our 1PN-accurate angular solution, given by Eq.~(\ref{phi_1PNa}).  

It is possible to correct this undesirable feature by introducing a certain  
PN accurate true anomaly 
$ \nu = 2 \arctan \biggl [ \biggl ( \frac{  e_{\phi} +1  }{ e_{\phi} -1 }
\biggr )^{1/2} \, \tanh \frac{u}{2} \biggr ]$, defined 
with the help of the {\it angular eccentricity} $e_{\phi}$.
This eccentricity parameter deviates from $e_r$ at PN orders by 
yet to be computed PN corrections.
It is easy to obtain the following 3PN 
accurate expression for $ \nu'$ in terms of $\nu$ 
\begin{align}
\nu' = &\,\nu+(f'-\frac{f'^2}{2}+\frac{f'^3}{4})\sin \nu\\\notag
&+(\frac{f'^2}{4}-\frac{f'^3}{4})\sin 2\,\nu
+\frac{f'^3}{12}\sin 3\,\nu \,,
\end{align}
where $f'$ should provide PN corrections connecting 
 $e_{\phi}$ and $e_r$.
We invoke the above relation in our $\phi- \phi_0$, given by Eq.~(\ref{Eq.2.29}),
and demand that there be no $\sin \nu'$ terms to 3PN 
order. 
The resulting 3PN-accurate parametric solution for the angular motion indeed 
reproduces Eq.~(\ref{Eq.2.29}) when restricted to 1PN order.
This procedure uniquely provides the PN corrections that connect $e_{\phi}$ to $e_r$, and 
 the resulting final parametrization for the angular
motion reads 
\begin{align}
\frac{2\pi}{\Phi}(\phi-\phi_0) = &\, \nu+(\frac{f_{4\phi}}{c^4}+\frac{f_{6\phi}}{c^6})\sin 2\,\nu+(\frac{g_{4\phi}}{c^4}+\frac{g_{6\phi}}{c^6})
\\\notag&\times\sin 3\,\nu
+\frac{h_{6\phi}}{c^6}\sin 4\,\nu+\frac{i_{6\phi}}{c^6}\sin 5\,\nu\,.
\end{align}

 We are now in a position to re-parametrize our 3PN accurate expression
 for $ t-t_0$, given by Eq.~(\ref{temp_t}), in terms of $ \nu$ with a procedure similar to the one outlined above.
This also ensures that we recover our Keplerian-type parametric expression 
for $l(u)$ at 1PN order. The improved 
 expression for the 3PN-accurate Kepler
equation reads
\begin{align}
\frac{2\pi}{P}(t-t_0) =& e_t\sinh u -u+(\frac{f_{4t}}{c^4}+\frac{f_{6t}}{c^6})\, \nu+(\frac{g_{4t}}{c^4}\\\notag&
+\frac{g_{6t}}{c^6})\sin \nu+\frac{h_{6t}}{c^6}\sin 2\,\nu+\frac{i_{6t}}{c^6}\sin 3\,\nu\,.
\end{align}
We observe that the transformation from $\nu'$ to $\nu$ ensures that the coefficients of $\nu$ terms appear only at the 2PN 
and 3PN orders.

 Collecting various results, we display in full 
 the third post-Newtonian
accurate 
generalized quasi-Keplerian parametrization for compact binaries 
in hyperbolic orbits as 
\begin{widetext}
\bes
\label{eq_3PN_GQKP}
\bea
r &=& a_r \left ( e_r\,\cosh u  -1 \right )\,,\\
\frac{2\pi}{P}(t-t_0) & =& e_t\sinh u -u+(\frac{f_{4t}}{c^4}+\frac{f_{6t}}{c^6})\, \nu+(\frac{g_{4t}}{c^4}+\frac{g_{6t}}{c^6})\sin \nu+\frac{h_{6t}}{c^6}\sin 2\,\nu+\frac{i_{6t}}{c^6}\sin 3\,\nu\,, \\
\frac{2\pi}{\Phi}(\phi-\phi_0) &=& \nu+(\frac{f_{4\phi}}{c^4}+\frac{f_{6\phi}}{c^6})\sin 2\,\nu+(\frac{g_{4\phi}}{c^4}+\frac{g_{6\phi}}{c^6})
\sin 3\,\nu+\frac{h_{6\phi}}{c^6}\sin 4\,\nu+\frac{i_{6\phi}}{c^6}\sin 5\,\nu\,,
\eea
\ens 
\end{widetext} 
where $ \nu  = 2\,\tan^{-1} \biggl [ \biggl ( \frac{e_{\phi} +1 }{e_{\phi} -1 } \biggr )^{1/2} \, \tanh \frac{u}{2} \biggr ]$.
Note that the 3PN-accurate expressions for the orbital elements $a_r, e_r^2, P=2\pi/n, e_t^2 , \Phi,$
and $e_{\phi}^2$ and the orbital
functions 
$  g_{4t},  g_{6t}, f_{4t}, f_{6t}, i_{6t}, h_{6t},
 f_{4\phi}, f_{6\phi}, g_{4\phi}, g_{6\phi},   i_{6\phi}, $
 and $h_{6\phi}$ are functions  of $E, h$ and $\eta$.
Their 3PN-accurate expressions in the modified harmonic 
coordinates arise from Eqs.~(A3) and (A4) of Ref.~\cite{MGS} 
and are given by
\begin{widetext}
\bes
\label{e:CoeffKP}
\setlength{\jot}{10pt}
\bea
a_r &=&  \frac{1}{{(2\,E)}}\bigg\{ 1+\frac{ (2\, E )}{4\,c^2} \left( 7-\eta \right) +
\frac{{{ (2\, E) }}^{2}}{16 c^4}\,\bigg[ 
(1+{\eta}^{2})
 \no&&
 +\frac {1}{(2\,E\,h^2)}
(64-112\,\eta)
\bigg] 
+{\frac {{{ (2\,E) }}^{3}}{192\,c^6}}\, 
\biggl [ 
-3+3\,\eta
-3\,{\eta}^{3}+\frac{1}{(2\,E\,h^2)}
\biggl (
768+ \left(123\,{\pi}^{2}-\frac{215408}{35}\right) \eta+1344\,{\eta}^
{2}\biggr)
\no 
&&
+
\frac{1}
{ (2\,E\, h^2)^2 }
\biggl (
6144+\left( -\frac{704096}{35}+492\,{\pi}^{2} \right) \eta
+1728\,{\eta}^{2}\biggr )
\biggr ]   \bigg\}\,,
\\
{e_{{r}}}^{2} &=&
1 + 2\,E\,h^2 +
\frac{(2\,E)}{4\,c^2}
\biggl \{ -24 +4
\,\eta+5\,\left(-3+ \eta \right) {(2\,E\,h^2)} 
\biggr \}
\no
&&
+ \frac{ (2\,E)^2}{8\,c^4}
\biggl \{
60+148\,\eta+2\,{\eta}^{2}
+\left( 80-45\,\eta+4\,{\eta}^{2} \right) {(2\,E\,h^2)}
\no&&
+\frac {8}{ (2\,E\,h^2)}
\left ( -16+28\, \eta \right )
\biggr \}
+ \frac{ (2\,E)^3 }{ 6720\,c^6} \biggl \{
2\,(1680-(90632+4305 \pi^2)\eta +33600\eta^2)\no&&
+4\,{\eta}^{3}
\bigg) \,
-\frac{80}{ (2\,E\, h^2)}\,\bigg(
1008+(-21130+861\pi^2)\,\eta+2268\,\eta^2\bigg)
\no
&&
-\frac{16}{(2\,E\,h^2)^2}
\biggl (
(53760 + (-176024 + 4305 \pi^2) \,\eta + 15120\,\eta^2)
\biggr )
\biggr \}\,,
\\
n&=&{{(2\,E)}}^{3/2} \bigg\{ 1-{\frac {{(2\,E)}} {8\,{c}^{2}}}\, 
\left( -15+\eta \right)
+{\frac {{{(2\,E)}}^{2} }{128{c}^{4}}} 
\biggl [ 555 
+30\,\eta
+11\,{ \eta}^{2}
\biggr ]\no&&
+
{\frac {{{(2\,E)}}^{3}}{1024\,{c}^{6}}}
\biggl [  653+111\eta+7\eta^2+3\eta^3 
\biggr ] 
\bigg\} \,,
\\
{\it e_t}^2 &=&1+{2\,E}\,{h}^{2}+
{\frac {{(2\,E)}}{4\,{c}^{2}}}\, {\bigg\{8-8\,
\eta
+ \left( 17-7\,\eta \right) {(2\,E\, h^2)} \bigg\} }
\no&&
+
\frac{{{(2\,E)}}^{2}}{8\,{c}^{4}} \bigg\{4\,(3+18\,\eta +5\,{\eta}^{2})
+  {(2\,E\,h^2)}( 112-47\,\eta
+16\,{\eta}^{2} )
\no&&
+\frac{16}{(2\,E\,h^2)} \left( -4 + 7\,\eta \right )
\bigg\}
+{
\frac {{{(2\,E)}}^{3}}{840\,c^{6}}}
\bigg\{ -70(42-830\,\eta+321\,\eta^2+30\,\eta^3)
\no&&-\frac{525}{8} (2Eh^2) (-528 + 200\, \eta - 77 \,\eta^2 + 24\, \eta^3) 
\no&&- \frac{3}{4(2\,E\,h^2)}
 \bigg(
73920 + (-260272 + 4305 \pi^2)\, \eta + 61040\, \eta^2\bigg)
\no&&
- \frac{1}{ (2\,E\,h^2)^2}
\bigg(
53760 + (-176024 + 4305 \pi^2)\, \eta + 15120\, \eta^2
\bigg)
 \bigg\}\,,
\\
f_{{4\,t}} &=&\frac{3\,(2\,E)^{2}}{2}\,\biggl \{ \frac{5 -2\,\eta }{ \sqrt{ (2\,E\,h^2)}}
 \biggr \}\,, 
\\
f_{{6\,t}} &=&{\frac {{{(2\,E)}}^{3}}{192(2Eh^2)^\frac{3}{2}}}
\biggl \{
\bigg(
10080+123\,\eta\,{\pi }^{2}-13952\,\eta
\no&&
+1440\,{\eta}^{2}
\bigg)
+(2Eh^2)36
\left (95-55\,\eta+18\,\eta^2
\right )
\biggr \}\,,
\\ 
g_{{4\,t}} &=& -\frac{1}{8}\,
\frac{ (2\,E)^2}{ \sqrt{ (2\,E\,h^2)}}
\biggl \{ (-15+ \eta)\,\eta \, \sqrt{(1 +2\,E\,h^2)}
\biggr \}\,,
\\
g_{{6\,t}}&=&{\frac {{{(2\,E)}}^{3}}{2240(2Eh^2)^\frac{3}{2}\sqrt{1+2Eh^2}}}
\bigg\{
35(2E h^2)^2 \eta \left(23\eta^2-175\eta+297\right)\no&&+(2E h^2)  \bigg(22400  + (49321 - 1435 \pi^2) \eta - 27300 \eta^2 + 
 1225\eta^3\bigg)\no
&&+385\, \eta^3-20965 \eta^2+\left(-1435 \pi ^2+43651\right) \eta+22400
\bigg\}\,,
\\
h_{{6\,t}} &=&\frac{{{(2\,E)}}^{3}}{16}\,\eta\,
\biggl \{ 
\frac{(1 +2\,E\,h^2)}{ (2\,E\,h^2)^{3/2} }
\left(116-49\eta+3\eta^2\right) 
\biggr \}\,,
\\
i_{{6\,t}} &=&
{\frac {\,{{(2\,E)}}^{3}}{192}}
\eta^3
\biggl ( 
\frac{  1 + 2\,E\,h^2 }{ 2\,E\,h^2}  
\biggr )^{3/2}\,\big(23-73\eta+13\eta^2\big),
\\
\Phi&=&2\,\pi \, \bigg\{ 1+{\frac {3}{{c}^{2}{h}^{2}}}+
-\frac{3(2E)^2}{4(2Eh^2)^2c^4}
\biggl [  -35+10\eta+(2Eh^2)\big(-5+2\eta\big)
\biggr ] 
\no&&
+{\frac {\,{{(2\,E)}}^{3}}{128\,{c}^{6}(2Eh^2)^3}}
\biggl [  
36960 +(615\pi^2-40000)\eta+1680\eta^2\no&&+  (2Eh^2)(10080+123\eta\pi^2-13952\eta+1440\eta^2)\no&&  + (2Eh^2)^2\big(120-120\eta+96\eta^2\big)
\biggr ] 
 \bigg\}\,, 
\\
f_{{4\,\phi}} &=&
\frac{{{(2\,E)}}^{2}}{8}
\,\frac{( 1+2\,E\,h^2)}{(2\,E\,h^2)^2}\,
 \, (1+19\,\eta-3\,\eta^2)\,,
\\
f_{{6\,\phi}}&=&{\frac {{{(2\,E)}}^{3}}{26880(2Eh^2)^3}}
\bigg\{
67200 + (994704 - 30135 \pi^2) \,\eta- 335160 \eta^2 - 4200 \eta^3 \no&&
+ 280 (2Eh^2)^2 (3 + 506\,\eta - 357 \,\eta^2 + 36\, \eta^3)\no&& + 
 (2Eh^2)\big(60480 + (991904 - 30135 \eta^2) \eta - 428400 \eta^2 + 8400 \eta^3\big)
\bigg\}\,,
\\
g_{{4\,\phi}} &=&
{\frac {(1-3\eta){{(2\,E)}}^{2}}{32}}
\frac{\,\eta\,}{(2\,E\,h^2)^2}
 ( 1 +2\,E\,h^2)^{3/2}\,,
\\
g_{{6\,\phi}}&=&
\frac{ (2\,E)^3}{768}\,\frac{ \sqrt{(1 +2\,E\,h^2)}}{(2Eh^2)^3}\,
\eta\bigg\{
36161 - 1435 \pi^2 - 28525\, \eta + 525\, \eta^2\no&& + 
 35(2Eh^2)^2 (14 - 49 \,\eta + 26 \,\eta^2) + 
 (2E h^2 ) (35706 - 1435 \pi^2 - 27510 \eta + 1750 \eta^2)
\bigg\}\,,
\\
h_{{6\,\phi}} &=&{\frac {{{(2\,E)}}^{3}}{192}}
\,\frac{{(1 +2\,E\,h^2)}^2}{(2\,E\,h^2)^3}\,\eta 
\left(82-57\,\eta+15\,\eta^2 \right)\,,
\\
i_{{6\,\phi}} &=&
\frac{\, (2\,E)^3}{256}\, \eta\frac{1-5\eta+5\eta^2}{ (2\,E\,h^2)^3}
\, (1 +2\,E\,h^2)^{5/2} \,,
\\   
{e_{{\phi}}}^{2}&=&
1 + 2\,E\,h^2 +
{\frac {{(2\,E)} }{4\,{c}^{2}}} \bigg\{ 
-24+ \left( -15+\eta \right) {(2\,E\,h^2)} \bigg\}
\no&&
+\frac{{{(2\,E)}}^{2} }{16\,{c}^{4}(2Eh^2)}
\bigg\{-416 +91\eta+15\eta^2+2(2Eh^2)\big(-20+17\eta+9\eta^2\big) +(2Eh^2)^2\big(160-31\eta+3\eta^2\big)\bigg\}
\no
&&
-{\frac {{{ (2\,E)}}^{3}}{13440\,{c}^{6}(2Eh^2)^2}}
\bigg\{2956800 + (-5627206 + 81795 \pi^2)\, \eta - 14490 \eta^2 - 7350 \eta^3\no&&
 - 
 (2Eh^2)^2 (584640 + (17482 + 4305 \pi^2)\, \eta + 7350\, \eta^2 - 
    8190 \eta^3)\no&& + 420 (2Eh^2)^3  (744 - 248\,\eta + 31 \,\eta^2 + 3\, \eta^3)\no&& + 
 14 (2Eh^2) (36960 + 7 (-48716 + 615 \pi^2)\, \eta - 225\, \eta^2 + 150\, \eta^3)\biggr \}\,.
\eea
\ens 
\end{widetext}
Let us recall that both the radial and temporal coordinates are scaled by $G\,m$,
and that the expressions for $a_r$ and $n$ are therefore given by 
 $a_r= 1/(2\,E)$ and $n= (2\,E)^{3/2}$ at the Newtonian order.
The three eccentricities
$e_r, e_t$ and $e_\phi$, which differ from each other
from the first post-Newtonian order, are related by
\begin{widetext}
\bes
\bea
{e_t}&=&{e_r}\, \bigg\{1+ {\frac {{(2E)}}{2{c}^
{2}}}(8-3\eta)+  {\frac {{{(2E)}}^{2}}{{c}
^{4}}} {\frac {1}{{(2E{h}^{2})}}}
\bigg[8-14\,\eta
+
\left(36 -19\,\eta+6\,{\eta}^{2} \right){(E {h}^{2})}
\bigg]
\no&&
+{\frac {{{(2E)}}^{3}}{3360{c}^{6}}}
{\frac {1}{{{(2E{h}^{2})}}^{2} }}
\bigg[
-420\,(2 \,E \,h^2)^2\, \left(10 \eta^3-34 \eta^2+65 \eta-160\right)+E\,h^2  \big(105840 \eta^2\\\notag&&+\left(4305 \pi ^2-354848\right) \eta+87360\big)+30240 \eta^2+\left(8610 \pi ^2-352048\right) \eta+107520\bigg] \bigg\}
\,,\\
{ e_\phi}&=&{ e_r} \,\bigg\{ 1-  {\frac {{ (2E)}}{2{c}^
{2}}} \eta-  {\frac {{{ (2E)}}^{2}}{32{c}^{4}}
}  {\frac {1}{{ (2E{h}^{2})}}} \bigg[160+357\eta-15{
\eta}^{2}
-\eta( -1+11{\eta}) { (2E{h}^{2})}\bigg]\\\notag&&
+  {\frac {{{ (2E)}}^{3}}{8960{c}^{6}
}}   {\frac {1}{{{ (2E{h}^{2})}}^{2}}} \bigg[
-70 (2 E h^2)^2 \,\eta\, \left(31 \eta^2-\eta-1\right)+5\,( 2 E h^2) \big(-1050 \,\eta^3\\\notag&&
+31304 \,\eta^2+\left(1435 \pi ^2-36546\right) \,\eta+4928\big)+2450\, \eta^3+166110 \,\eta^2+\big(18655 \pi ^2\\\notag
&&-1854\big) \,\eta-412160
\bigg] \bigg\}\,.
\eea
\ens
\end{widetext}
These relations allow one to choose a specific eccentricity parameter 
to describe a PN accurate hyperbolic orbit.

  Following the above detailed procedure, it is straightforward to obtain 3PN-accurate 
expressions for the above listed quantities also in an ADM-type 
gauge. The 3PN-accurate Keplerian-type parametric solution arises 
from Eqs.~(A1) and (A2) of Ref.~\cite{MGS} and is structurally identical to 
Eqs.~(\ref{eq_3PN_GQKP}). This is expected
as Eqs.~(A1), (A2) and (A3) and (A4) of  Ref.~\cite{MGS} are 
polynomials of the same degree though their coefficients are different.
The ADM versions of Eqs.~(\ref{e:CoeffKP}) are listed 
in Appendix~\ref{AppA}.

  We are now in a position to explore the possibility of obtaining 
our 3PN-accurate hyperbolic solution from its eccentric counterpart through analytic continuation.
A close inspection of our results reveals that the 3PN-accurate expression for the orbital element $n$ in Eqs.~(\ref{e:CoeffKP}),
is structurally different from 
its eccentric counterpart, given by
Eq.~(25c) of Ref.~\cite{MGS}.
Moreover, the structure of the relevant two Kepler equations is 
different (compare Eq.~(19b) of  Ref.~\cite{MGS} with our Eq.~(\ref{eq_3PN_GQKP}b)).
Therefore, it is reasonable to expect that additional arguments 
may be required to obtain {\it practically viable} analytic continuation arguments 
for extracting our main results from that 
of Ref.~\cite{MGS}.
We begin from the eccentric Kepler equation, given by Eq.~(24b) of  Ref.~\cite{MGS},
which may be written as
\begin{widetext}
\bea
\label{KE_e}
l=\frac{2\pi}{P_e}\left ( t - t_0 \right ) &=&
u' -e_t\,\sin u' +
\left ( \frac{g'_{4t}}{c^4} + \frac{g'_{6t}}{c^6} \right )\,(\nu' -u')
 +
  \left ( \frac{f'_{4t}}{c^4} + \frac{f'_{6t}}{c^6} \right )\,\sin \nu'
  + \frac{i'_{6t}}{c^6} \, \sin 2\,\nu'
  + \frac{h'_{6t}}{c^6} \, \sin 3\, \nu' \,\no \,,
\eea
\end{widetext}
where {\it primed} variables denote an eccentric binary 
and $P_e$ stands for the 3PN-accurate orbital period 
of an eccentric binary.
The presence of the term
$ \nu'-u' \equiv 2\tan^{-1}\left(\beta'_{\phi}\sin u'/(1-\beta'_{\phi}\cos u')\right)$ in the above Kepler equation,
where $\beta_{\phi}=(1-\sqrt{1-e_{\phi}^2})/e_{\phi}$,
leads to certain imaginary terms 
while adapting the usual argument of analytic continuation, namely $ u' \rightarrow \imath v $ and 
 $\sqrt{-E} \rightarrow \imath\,\sqrt{E}$, 
to obtain its hyperbolic version \cite{DD1985}.

   At 1PN order, the above arguments ensure that the expression for $P_e$ becomes a purely imaginary quantity, i.e., $\imath P_\text{hyp}$
   and that $u'- e_t\, \sin u'$ becomes $ \imath v-\text{e}_t\,\sin(\imath v)$.
This guarantees that $ (P_e)  (u'- e_t\, \sin u')/(2\, \pi) $ leads to   $ (P_\text{hyp})  (e_t\, \sinh v -v)/(2\, \pi) $.
This observation influenced us to consider an expression for $(t- t_0)$, as given by 
 Eq.~(\ref{KE_e}), rather than $l = n(t-t0)$, while invoking the usual arguments for analytic continuation (AAC) .
It is easy to show that the 3PN-accurate eccentric expression for $n$
 gives a complex quantity rather than a purely imaginary one under the AAC. This is essentially due to 
 the presence of $(-2\,E\, h^2)$ terms present in Eq.~(25c) in Ref.~\cite{MGS}. 
Similar arguments also apply to the terms $ u'- e_t\, \sin u'+ ( \frac{g'_{4t}}{c^4} + \frac{g'_{6t}}{c^6}  )\,(\nu' -u')$ 
in the 3PN-accurate eccentric Kepler Equation under the AAC.
However, the product of $ P_e/2\,\pi$ and the above terms while substituting
  $ u' \rightarrow \imath v$ and  $\sqrt{-E} \rightarrow \imath\,\sqrt{E}$ becomes 
a real quantity
and can be identified with $ (P/2\,\pi) \times [ e_t\, \sinh u -u + ( \frac{f_{4t}}{c^4} + \frac{f_{6t}}{c^6}  )\, \nu ]$. 
Here, $ P/(2\,\pi)$ is the PN-accurate inverse of $n$, given by our Eqs.~(\ref{e:CoeffKP}).
It should be noted that this procedure ensures that the complex quantities that we encountered while applying the AAC
in Eq.~(\ref{KE_e}) are now properly handled to obtain our  3PN-accurate hyperbolic solution.
Let us emphasize that we were able to formulate this reasoning mainly because of the availability 
of our 3PN-accurate hyperbolic solution, obtained from our detailed ab-initio computations.
In other words, it is rather difficult to extract a 3PN-accurate orbital element $n$ for hyperbolic orbits from its 
eccentric version simply by invoking the arguments for analytic continuation of Ref.~\cite{DD1985}.
We require re-definitions of certain terms 
in the eccentric Kepler equation to obtain its hyperbolic version through analytic continuation. These re-definitions, however, can be worked 
out if the actual hyperbolic solution is available, computed from first principles as done in this paper.
Finally, we observe that all other eccentric orbital elements and functions transition smoothly into their 
hyperbolic counterparts while employing the AAC.
The extraction of our 3PN-accurate hyperbolic solution from its eccentric counterpart, as noted earlier,
provides an additional test for  the correctness of the lengthy expressions present in Ref.~\cite{MGS}.

 We have also adapted for our purposes a consistency check which  was devised in Ref.~\cite{MGS} to test the 
 fidelity of the 3PN-accurate eccentric parametrization and 
 its PN-accurate orbital elements and functions.
The idea is to compute 3PN-accurate expressions for 
 $\dot r^2 $ and $\dot \phi^2$ using our parametric solution, via
 $\dot r^2 = \left ( \frac{d r}{du}\, \frac{du}{dt} \right )^2$
 and ${\dot \phi}^2 = {\left ( \frac{d \phi}{dv}\, \frac{d v}{du}\, \frac{du}{dt}
\right )}^2 $.
These lengthy 3PN-accurate expressions are first obtained in terms of $E, h, \eta$ and $( e_r\, \cosh u -1)$
and are later converted in terms of $E, h, \eta$ and $r$ while using 
our 3PN-accurate expression for $r= a_r(E,h,\eta) \left( e_r\, \cos hu -1 \right )$.
A detailed check is provided by comparing these parametric expressions for $\dot r^2 $ and $\dot \phi^2$ with those extracted 
from Eqs.~(A3) and (A4) in Ref.~\cite{MGS}.
Note that these equations  arise from the 3PN-accurate expressions for the 
orbital energy and angular momentum
as evident by examining Eqs.~(22)  and (23) 
and the associated discussions 
in Ref.~\cite{MGS}.
We have verified that the above two sets of 3PN-accurate expressions for 
$\dot r^2 $ and $\dot \phi^2$ in terms $E,h,\eta$ and $r$ are identical to each other
in the case of hyperbolic orbits.
Let us emphasize that this check is very sensitive to the structure
of the parametric solution and the explicit 
PN-accurate expressions for the various orbital elements and functions.
Therefore, the complete agreement to 3PN order between the parametric and Hamiltonian-based sets of 
 $\dot r^2 $ and $\dot \phi^2$ expressions -- along with our 
improved analytic continuation relations -- provide 
powerful checks on our 
3PN-accurate generalized quasi-Keplerian parametrization for 
compact binaries in hyperbolic orbits.
Additionally, we have verified  that our results are in agreement 
with Ref.~\cite{DD1985} at 1PN order.
In what follows, we apply our 3PN-accurate Keplerian type parametric solution to obtain time-domain
gravitational waveforms for compact binaries in hyperbolic orbits while incorporating effects of 
GW emission.

\section{ GW polarization states for compact binaries in 3.5PN-accurate hyperbolic orbits}\label{sec:hcp3.5PN}
\label{sec:II}

 This section presents an efficient prescription to obtain temporally evolving GW polarization states for compact 
 binaries moving in fully 3.5PN-accurate hyperbolic orbits.
Clearly, this requires us to prescribe a way of incorporating the dissipative effects of GW emission appearing at 2.5PN and 3.5PN orders into our 
3PN-accurate orbital dynamics.
 With the help of Refs.~\cite{DGI,KG06,LGGJ}, this is pursued in steps which we will briefly outline below.
We begin by considering the following expressions for 
the quadrupolar (or Newtonian) order GW polarization states, 
$h_+|_Q $ and $h_{\times}|_Q $, for compact binaries in non-circular orbits, available in Ref.~\cite{LGGJ}, which read
\begin{subequations}
  \label{11}
  \begin{align}
    h_+|_{\rm Q} = &-\frac{G\, m\, \eta }{c^4\,R}\;\times  \label{9}\\
		   &\bigg\{\left(1+C_{\theta}^2\right)\bigg[\left(z+r^2\dot{\phi}^2-\dot{r}^2\right)\cos2 \phi\;\nonumber \\
		   &+2r\dot{r}\dot{\phi}\sin 2\phi\bigg]+S_{\theta}^2\left(z-r^2\dot{\phi}^2-\dot{r}^2\right)\bigg\}~, \\
    h_{\times}|_{\rm Q} = &-2\frac{G\, m\, \eta }{c^4\,R}C_{\theta}\;\times   \label{8} \\
			  &\bigg\{\left(z+r^2\dot{\phi^2}-\dot{r}^2\right)\sin 2\phi-2r\dot{r}\dot{\phi}\cos 2\phi\bigg\}\,.
  \end{align}
\end{subequations}
The parameter $z$ is related to the radial coordinate of the orbit by $z= G\,m/r$, while 
 $R$ is the radial distance to the source, and $C_{\theta}= \cos \theta$, $S_{\theta}= \sin \theta$ with 
$\theta$ being the orbital inclination.
Obviously,  the temporal evolutions of $h_+|_Q(t) $ and $h_{\times}|_Q(t) $ require  
a prescription for evolving $r, \dot r =  \text{d}r/\text{d}t, \phi$ and $\dot \phi = \text{d}\phi/\text{d}t$ in time.

In the next step, we obtain fully 3PN-accurate parametric  expressions for the dynamical variables appearing in 
the above expressions for $h_+|_Q(t) $ and $h_{\times}|_Q(t) $.
This requires parametric expressions not only for $r$ and $\phi$, available in the previous section, but
also for $\dot r $ and  $\dot \phi$. We obtain 3PN-accurate parametric expressions for  $\dot r $ and $\dot \phi$ by noting
that $ \dot r =  ( dr/du) \times ( du/dt) $ and $\dot \phi = ( d \phi/d \nu ) \times ( d \nu/du ) \times ( d u/dt) $.
These expressions are  provided in terms of a certain gauge-invariant dimensionless PN expansion parameter 
$\xi = \frac{G\,m\,n}{c^3}$, where $n=\frac{2\pi}{P}$ as defined in Eq.~(\ref{e:CoeffKP}c), the {\it time eccentricity} $e_t$
and the eccentric anomaly $u$.
The dynamical variables have to derived carefully, as we introduced
scaled coordinates in the previous section.
Our particular choice of variables is influenced by the ease with which we can specify various initial conditions during the numerical
construction of GW templates. 
To obtain 3PN-accurate temporal evolutions of $r, \dot r, \phi$ and $\dot \phi$, we also need to express the right-hand side of the 3PN-accurate  
Kepler equation in terms of $\xi$ and $e_t$. 

The third step involves including
the effects of GW emission during hyperbolic passages. This is accomplished by providing differential equations for 
$d \xi/dt$ and $ de_t/dt$, whose derivation is influenced by Refs.~\cite{DGI,KG06}.
These equations, as expected, incorporate   
radiation reaction effects entering the orbital dynamics at 2.5PN and 3.5PN orders.
Through a numerical solution of the Kepler equation along with these two coupled differential equations for $\xi$ and $e_t$,  
 we obtain the fully 3.5PN-accurate temporal evolution for  $r, \dot r, \phi$ and $\dot \phi$.
This enables us to construct $h_+|_Q(t) $ and $h_{\times}|_Q(t) $ for compact binaries in 3.5PN-accurate 
hyperbolic orbits.
Finally, we provide a 3PN accurate expression for $\xi$ in terms of a certain PN-accurate gauge-dependent impact parameter $b$ and 
time eccentricity $e_t$, as it is very convenient to characterize hyperbolic orbits through their impact parameters and eccentricities.
Thus, we obtain ready-to-use GW templates for compact binaries in PN-accurate hyperbolic orbits.

 In the following, we provide explicit expressions for various dynamical variables in terms of $\xi, e_t$ and $u$ that will be required
 for obtaining $h_+|_Q(t) $ and $h_{\times}|_Q(t) $. For the sake of readability, we will only explicitly list the  
 1PN-accurate expressions for these dynamical variables, given  
 in terms of $\xi, e_t$ and $u$ as 
\onecolumngrid
\begin{subequations}
  \label{eq:quasi_kepl_nonsp}
  \begin{align}
    r (u) =\; &\frac{Gm}{c^2}\frac{1}{ \xi^{2/3}}(e_{\rm t}\cosh u-1)\,\left\{1+ \xi^{2/3}\;\frac{2\eta-18-(6-7\eta)e_{\rm t}\cosh u}{6\left(e_{\rm t}\cosh u-1\right)}\right\}~, \label{1}\\
    \dot{r} (u) =\; & \xi^{1/3}\frac{c\, e_{\rm t}\sinh u}{e_{\rm t}\cosh u-1}\left\{1- \xi^{2/3}\frac{6-7\eta}{6}\right\}~,\\
    \phi (u) - \phi_0 =\; &2\arctan \left[\left(\frac{e_{\phi}+1}{e_{\phi}-1}\right)^{1/2}\tanh u/2\right]\,\left\{1+ \xi^{2/3}\frac{3}{e_{\rm t}^2-1}\right\}~,\\ 
    \dot{\phi} (u) =\; &\frac{ n\, \sqrt{e_{\rm t}^2-1}}{\left(e_{\rm t}\cosh u-1\right)^2}\bigg\{1- \xi^{2/3}\,\frac{\left[3-\left(4-\eta\right)e_{\rm t}^2+\left(1-\eta\right)e_{\rm t}\cosh u\right]}{\left(e_{\rm t}^2-1\right)\left(e_{\rm t}\cosh u-1\right)}\bigg\}\label{4}~. 
  \end{align}
\end{subequations}
The lengthy 3PN-accurate versions of these expressions are provided  
 in Appendix~\ref{AppB}.
\twocolumngrid 

 It should be obvious that temporal evolutions for the 3PN version of above equations, namely Eqs.~(\ref{eq:quasi_kepl_nonsp}), 
 require a 3PN-accurate Kepler equation in terms of $\xi$ and $e_t$ that connects $l$ and $u$.
This 3PN-accurate equation in MH gauge is given by   
\begin{widetext}
\bes
\label{meanAnomaly}
\bea
\setlength{\jot}{9pt}
l&=&n(t-t_0)=l_\text{N}+l_\text{1PN}+l_\text{2PN}+l_\text{3PN}\,,\\
l_\text{N}&=&{e_t} \sinh u-u\,,\\
l_\text{1PN}&=&0\,,\\
l_\text{2PN}&=&\frac{\xi^{4/3} }{8 \sqrt{e_t^2-1}}\Big[12 \nu  (5-2 \eta)-e_t (\eta-15) \eta \sin \nu\Big]\,,\\
l_\text{3PN}&=&\frac{\xi^2 }{6720 (e_t^2-1)^{3/2} (e_t \cosh u-1)}\Bigg\{35 \nu  \Big\{96 e_t^2 \Big[\eta (11 \eta-29)+30\Big]+\eta (960 \eta+123 \pi ^2-13184)+8640\Big\}\noo
\times(e_t \cosh u-1)+840 e_t \sqrt{e_t^2-1} (\eta-4) \sinh u \Big[e_t (\eta-15) \eta \cos \nu+24 \eta-60\Big]+e_t \sin \nu\, (e_t \cosh u-1)\,\noo
\times \Bigg[\eta \Big\{70 e_t^2 \Big[\eta (39 \eta-239)+7\Big]-4 \Big[70 \eta (\eta+222)-35967\Big]-4305 \pi ^2\Big\}+70 e_t \eta \Big\{e_t \Big[\eta (13 \eta-73)+23\Big] \noo\times\cos 2\nu+12 \Big[\eta (3 \eta-49)+116\Big] \cos \nu\Big\}+67200\Bigg]\Bigg\}\,.
\eea
\ens
\end{widetext}
 The above equation allows us to adapt Mikkola's method, developed to  
numerically solve the classical Kepler equation for hyperbolic orbits as detailed in Sec. 4 of Ref.~\cite{Mikkola87}. 
Mikkola's very effecient and computationally inexpensive approach approximates the classical Kepler equation as a cubic polynomial 
in an auxiliary variable $s(u)$, finding its roots and substantially reducing the error of the initial guess through a fourth-order extension of Newton's method.
We employ Mikkola's procedure in an iterative 
manner to tackle PN corrections to the classic Kepler equation appearing at 2PN and 3PN orders.
It should be noted that our 3PN-accurate Kepler equation is identical to the classical (Newtonian) Kepler equation at 1PN order,
which is only possible due to the use of the time eccentricity $e_t$ as a parameter to specify the orbit.
To solve above 3PN-accurate Kepler equation, we tackle   
the 1PN-accurate Kepler equation, namely $ l = n (t-t_0) =  e_t\,\sinh u -u$, using Mikkola's original
prescription and obtain a 1PN-accurate expression for $u(l)$.\\
\indent This method requires us  to express $l$ in terms of 
a new variable
$s' = \sinh\frac{u}{3}$,
\begin{align}
l = e_t\, (3 s'+ 4 s'^3)-3\ln(s'+\sqrt{1+s'^2})\,,
\end{align}
and truncating it to the third order in $s'$, 
\begin{align}
l =3 (1-e_t) s' + (4e_t+\frac{1}{2})\,s'^3\,.
\end{align}
This third order polynomial can be solved in a closed form, say, $s'=s'(l\,;\,e_t)$. To minimize the error, replacing $s'$ to 
$$\omega(l): = s'(l)+\frac{0.071s'(l)^5}{(1+0.45 s'(l)^2)(1+4s'(l)^2)e_t}.$$ Now we can get the most accurate solution,
\begin{align}
u(l)\,= l - e_t\, (3 \omega(l)+ 4 \omega(l)^3)\,.
\end{align}
The accuracy of the solution can further improved by the use of Newton method as noted in Ref.~\cite{Mikkola87}. \\
\indent This allows us to express numerically the 2PN and 3PN corrections  that appear on the right-hand side of Eq.~(\ref{meanAnomaly}) in terms of $ (\,\xi, e_t, l\,)$.
We now introduce a certain parameter $l'$ such that $l'= l - l_{4,6}$, where 
$l_{4,6}$ are the 2PN and 3PN corrections present in Eq.~(\ref{meanAnomaly}) which are 
evaluated using 1PN-accurate $u(l)$.
The 3PN accurate $u(l)$ is obtained, as expected, by solving 
$ l' =  e_t\,\sinh u -u$, once again employing Mikkola's method.
In this way, we pursue an
accurate and efficient solution to our 3PN accurate Kepler equation
which allows us to compute the 3PN-accurate temporal evolutions for the dynamical variables present
in  our expressions for $h_+|_Q(t) $ and $h_{\times}|_Q(t) $.
We note, in passing, that to obtain these 3PN-accurate expressions for 
 $r, \dot r, \phi$ and $\dot \phi$, we have used unique 3PN-accurate expressions that provide 
 $2\,E$ and $h$ in terms of $\xi$ and $e_t$ by inverting the relevant expressions present 
 in our parametric solution. Further, we have also employed 3PN-accurate relations that provide
 $e_r$ and $e_{\phi}$ in terms of $e_t, \xi$ and $\eta$.
We are now in a position to discuss how GW emission effects are incorporated.
 
  GW emission influences binary dynamics at 2.5PN and 3.5PN orders, and we incorporate these effects 
 by adapting the phasing formalism developed for eccentric binaries (detailed in Ref.~\cite{DGI,KG06})
to hyperbolic encounters. 
This requires us to compute time derivatives of the 1PN-accurate
expressions for the conserved orbital energy and angular momentum of binaries
in non-circular orbits, given by Eqs.~(3.35) and (3.36) in Ref.~\cite{BS89}.
These time derivatives are obtained using 
PN-accurate equations of motion that include both conservative
and reactive terms to 1PN order, e.g., given by Eq.~(3.34) of 
Ref.~\cite{BS89} and  Eqs.~(28), (29) of  \cite{KG06}.
The resulting expressions for $dE/dt$ and $dh/dt$ are adapted for hyperbolic orbits 
with the help of our 1PN-accurate parametric expressions for the dynamical variables $r, \dot r$ and $ \dot \phi$,
expressed in terms of $n,e_t, u$.
Using our 1PN-accurate expressions for $n= 2\,\pi/P$ and $e_t^2$ in terms of 
the conserved orbital energy and angular momentum, $dE/dt$ and $dh/dt$ then lead to the desired equations for $dn/dt$ and 
$d e_t/dt$ in modified harmonic gauge: 
\bes
\begin{align}\label{dndt}
\frac{dn}{dt}=&\,\frac{8c^6\eta\,\xi^\frac{11}{3}}{5\,G^2\,m^2\,\beta^6}\,\Big[35(1-e_t^2)+49\beta+32\beta^2+6\beta^3-9\beta e_t^2\Big]\no
&+\frac{2c^6 \eta}{35 \beta^9}\xi^{\frac{13}{3}}\Big\{\beta^6 (180-588 \eta)+\beta^5 (1340-5852 \eta)\no
&+2 \beta^4 \Big[9 e_t^2 (21 \eta-1)-8589 \eta+1003\Big]+35 \beta^3 \Big[e_t^2\no&\times
 (244 \eta-5)-684 \eta+21\Big]+35 \beta^2 (e_t^2-1) \Big[9 e_t^2\no&
\times (2 \eta-17)+454 \eta+193\Big]-21 \beta (e_t^2-1)^2 (140 \eta\no
&+657)+5880 (e_t^2-1)^3\Big\}\,,
\end{align}

\begin{align}\label{detdt}
\frac{de_t}{dt}=&\frac{8c^3\eta(e_t^2-1)}{15\,G\,m\,\beta^6\,e_t}\xi^\frac{8}{3}\Big[35(1-e_t^2) +( 49-9e_t^2) \beta+ 17 \beta^2\no
& + 3 \beta^3 \Big]-\frac{2c^3\eta}{315\beta^9e_t}\xi^\frac{10}{3}\Big\{-17640(-1 + e_t^2)^4 + 63 \beta\no&\times
 (-1+ e_t^2)^3 (657 + 140 \eta)- 105 \beta^2 (-1 + e_t^2)^2 \Big[13\no&\times
 + 454 \eta + 9 e_t^2 (3 + 2\eta)\Big]- \beta^4 (-1 + e_t^2) \Big[36825\no&
 - 53060 \eta + 9 e_t^2 (-2169 + 560 \eta)\Big]+ 6 \beta^6 \Big[360 - 553 \eta \no&
+ e_t^2 (-444 + 637 \eta)\Big] - 28 \beta^3 (-1 + e_t^2) \Big[29 (63 - 95 \eta) \no&
+ e_t^2(-1767 + 1105 \eta)\Big]+\beta^5 \Big[10215 - 18088 \eta + e_t^2 \no&\times (-12735 + 20608 \eta)\Big]\Big\}\,.
\end{align}
\ens
where $\beta = e_t\cosh u-1$. 
We have verified that these expressions can also be obtained by the usual calculations based on balance arguments.
In this approach, one 
 differentiates our 1PN-accurate expressions for 
$n$ and $e_t$ while using  1PN-accurate expressions for the far-zone fluxes, given for example by 
Eqs.~(17) and (18) of Ref.~\cite{JS92}, to replace the time derivatives of the conserved energy and angular momentum variables.
The resulting expressions for $dn/dt$ and $de_t/dt$, adapted for 1PN-accurate hyperbolic orbits, were found to be identical
to Eqs.~(\ref{dndt}) and (\ref{detdt}).

It is rather convenient to characterize hyperbolic binaries in terms of an impact parameter $b$,
as these GW events are qualitatively similar to scattering processes.
We define a PN-accurate impact parameter $b$ 
such that $b\, \text{v}_{\infty} = | \bm r \times \textbf{ v} |$ when $ |\bm r | \rightarrow \infty$, while 
$\text{v}_{\infty}$ stands for the relative velocity at infinity \cite{BS89}. 
The explicit 3PN-accurate expression for $b$  in terms 
of $\xi$ and $e_t$ in modified harmonic gauge reads 
\begin{align}\label{b_hyp}
b=& \frac{Gm}{c^2}\frac{\sqrt{e_{\rm t}^2-1}}{ \xi^{2/3}}\Bigg\{\,1- \xi^{2/3}\left(\frac{\eta-1}{e_{\rm t}^2-1}+\frac{7\eta-6}{6}\right)\\\notag
&+\xi^{4/3}\bigg[1-\frac{7}{24}\eta+\frac{35}{72}\eta^2+\frac{3-16\eta}{2(e_t^2-1)}+\frac{7-12\eta-\eta^2}{2(e_t^2-1)^2}\bigg]\\\notag
&+\xi^{2}\,\bigg[-\frac{2}{3}+\frac{87}{16}\eta-\frac{437}{144}\eta^2+\frac{49}{1296}\eta^3+\\\notag
&+\frac{36-378\eta+140\eta^2+3\eta^3}{24(e_t^2-1)}+\frac{1}{6720(e_t^2-1)^2}\big\{248640\\\notag
&+(-880496+12915\pi^2)\,\eta+40880\,\eta^2+3920\,\eta^3\big\}\\\notag
&+\frac{1}{1680(e_t^2-1)^3}\big\{73080+(-228944+4305\pi^2)\eta\\\notag
&+47880\eta^2+840\eta^3\big\}\bigg]\Bigg\}~.
\end{align}

At 1PN order, we are in agreement with Ref.~\cite{LGGJ}.
 This variable is essentially invoked to allow
for an easy visualization of the trajectories of hyperbolic binaries.
\onecolumngrid

\begin{figure}[H]
	\centering
	\includegraphics[width=1\textwidth]{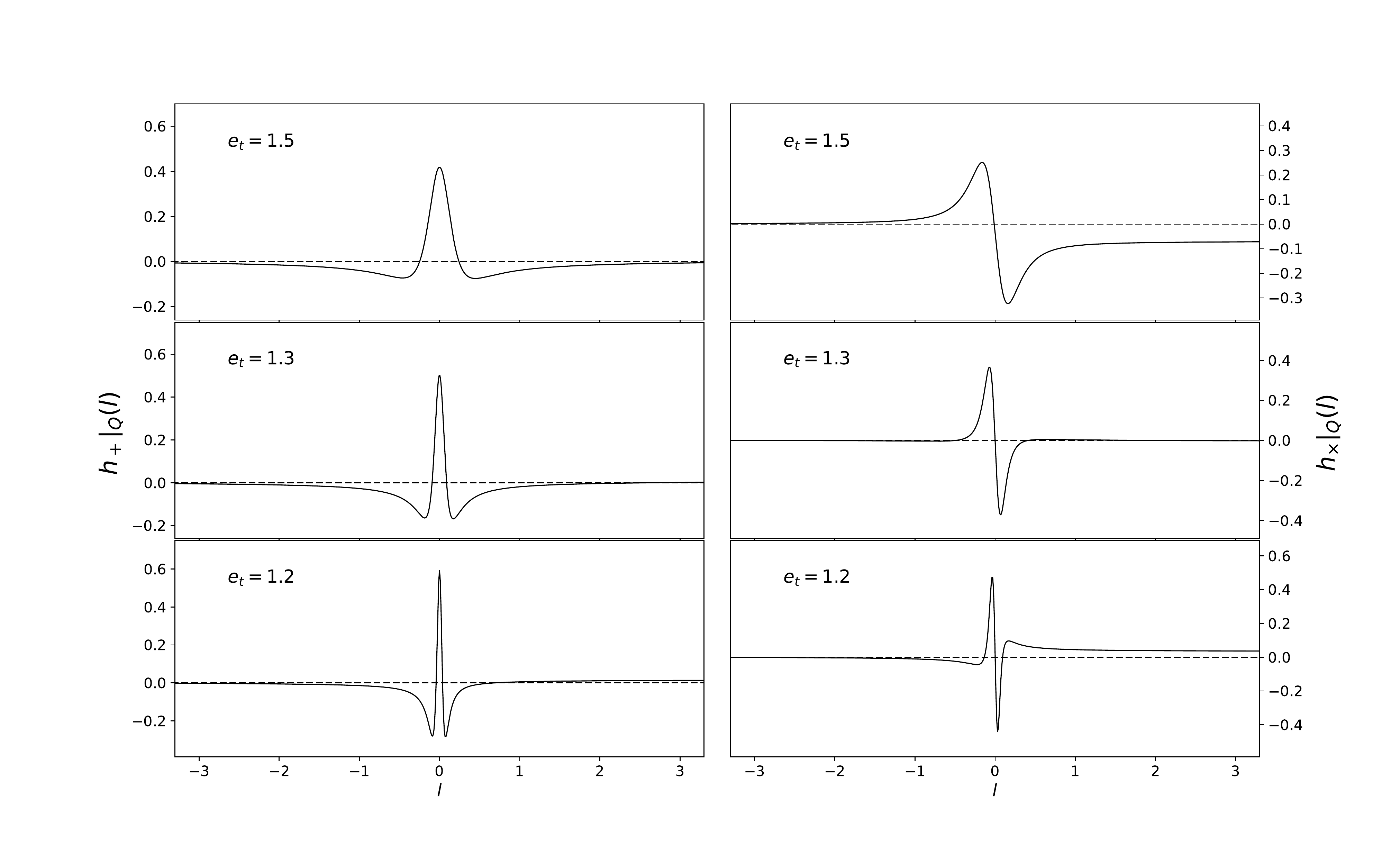}
\caption{Scaled $H_+|_Q(l) $ and $H_{\times}|_Q(l) $ plots for 
non-spinning compact binaries with total mass 
	$m= 20\, M_{\odot}$ and mass ratio $q=1$.
We let the eccentricity $e_{t} $ take three values $1.5$, $1.3$ and $1.2$, while
choosing an impact parameter $b \sim 30 \,Gm/c^2$ and inclination angle $\theta=\frac{\pi}{4}$. 
We observe the expected linear memory effect in the cross polarization state.
}
\label{fig:1}
\end{figure}
\twocolumngrid

 With above inputs, it is possible to obtain temporally evolving Newtonian (quadrupolar) GW polarization states, $h_+|_Q(t) $ and $h_{\times}|_Q(t) $,
 associated with compact binaries in 3.5PN-accurate hyperbolic orbits.
It is convenient to numerically solve a system of three coupled differential equations, namely $dn/dt, de_t/dt$ and $dl/dt =n$. 
The resulting values of parameters $n,e_t$ and $l$ at a given epoch are then employed to obtain a 3PN-accurate value for $u(l)$ 
by the application of Mikkola's method as described above. With a knowledge of $n,e_t, l$ and $u$, 
we can then evaluate 
our 3PN-accurate expressions for $r, \dot r, \phi$ and $\dot \phi$. Thus, we are able to numerically 
provide $h_+|_Q(t) $ and $h_{\times}|_Q(t) $
 from compact binaries in 3.5PN-accurate hyperbolic orbits. In the following, we discuss plots that
 demonstrate the approach and point out a feature of the waveforms previously not mentioned in the literature.

\begin{figure}[H]
	\centering
	\includegraphics[width=1\textwidth]{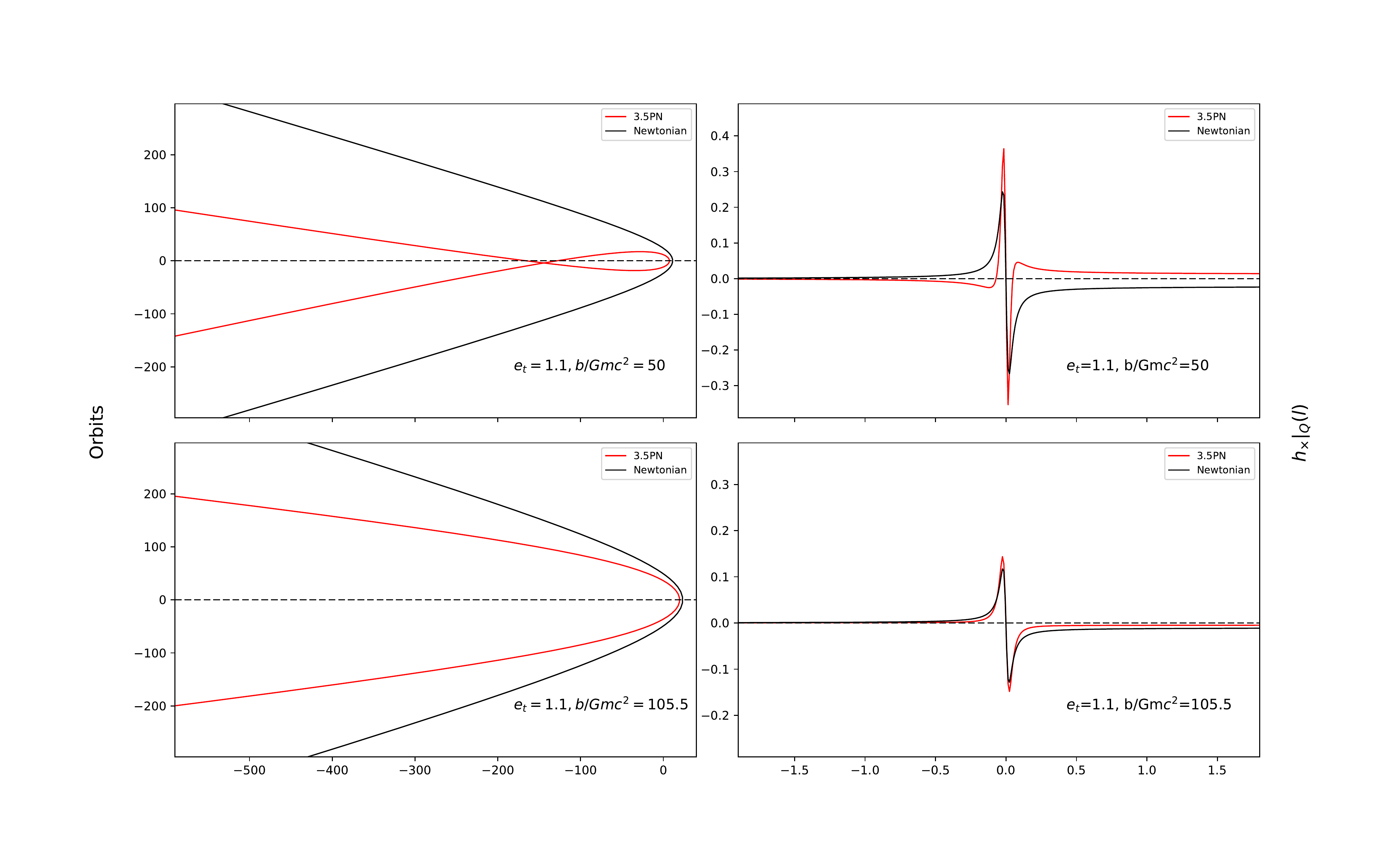}
\caption{Trajectories and the associated scaled $H_{\times}|_Q(l) $ 
for hyperbolic compact binaries, with a choice of two different impact parameters $b$, eccentricity $e_t=1.1$, total mass $m=20 M_{\odot}$,
 mass ratio $q=1$, and inclination angle $\theta=\frac{\pi}{4}$. For the trajectories, we adopt the geometric unit system. 
 Newtonian and 3.5PN-accurate hyperbolic orbits are denoted by black and red lines, respectively.
The orbital trajectory of the relativistic system  is clearly different, especially for hyperbolic 
passages with smaller $b$ values, which is attributed to the advance of periastron.
Relativistic effects also change the nature of the waveforms, as evident from the associated
$h_{\times}|_Q(l) $  plots. 
}
	\label{fig:2}
\end{figure}

In Fig.~\ref{fig:1}, we plot scaled quadrupolar GW polarization states, $H_+|_Q(l) $ and $H_{\times}|_Q(l) $,
for hyperbolic passages with $b \sim  30 \, G\,m/c^2$ for compact binaries 
having $m= 20\, M_{\odot}$ and $\eta =\frac{1}{4}\,(q=1)$, while allowing $e_t$ to take three different values.
Here, $H_+|_Q(l) $ and $H_{\times}|_Q(l) $ denotes waveforms that have been scaled by $ G\,m /c^2\,R$.
We observe, as
expected, the linear memory effect for the cross polarization \cite{LGGJ}.
We display in Fig.~\ref{fig:2} the trajectories of compact binaries under the influence of Newtonian 
 and fully 3.5PN-accurate orbital dynamics (respectively in black and red) and their associated $H_{\times}|_Q(l) $.
For these, we have chosen $e_t = 1.1$ while we let the impact parameter $b$ take two different values, namely, $ \sim 50\, G\, m/c^2 $ and $ \sim 106 \, G\,m/c^2$.
These particular $v$ values were chosen to highlight the effect of PN corrections compared to the familiar Newtonian hyperbolic orbit. 
We observe that the periastron advance forces the 3.5PN-accurate orbital trajectory to cross its earlier path,  
a feature which is absent in the Newtonian system.  Additionally,
this feature disappears for large impact parameter values. This is expected, as the periastron advance is small 
for configurations with a large impact parameter, which results in Newtonian-like trajectories.
We have also verified that the PN corrections in $\Phi/2\, \pi$ indeed converge to its 1PN value in above 
cases; this ensures that the crossing of the trajectory is a physical effect.
Interestingly, this PN effect leads to sharper GW polarization states, and it will be interesting to explore possible 
data analysis implications for such hyperbolic passages.

\onecolumngrid

\twocolumngrid

\section{Conclusions}\label{sec:conclusions}

  In this paper, we provided `ready-to-use` time-domain GW polarization templates for compact binaries moving in fully 3.5PN-accurate 
hyperbolic orbits. A crucial input for constructing these waveforms is our 
\textit{ab-initio} 
derivation of 3PN-accurate Keplerian type 
parametric solution for compact binaries in hyperbolic orbits.
Our effort extended the classic 1PN result of Damour and Deruelle,obtained by the argument of analytic continuation
to 3PN order \cite{DD1985}.
Additionally, we provided two critical checks to verify the correctness of our solution and its lengthy 3PN-accurate expressions.
We incorporated the effects of GW emission, occurring at 2.5PN and 3.5PN orders in the orbital dynamics, by adapting  for hyperbolic orbits
GW phasing formalism for eccentric inspirals, detailed in Refs.~\cite{DGI,KG06}.
This is how we constructed our PN-accurate GW templates, namely temporally evolving GW polarization states, for hyperbolic encounters.

 The present effort should be useful in a number of on-going investigations.
Our templates are being implemented in the LSC Algorithm Library Suite (LALSuite) \cite{aLIGO}.
This is to explore the possibility of searching for the presence of such GW events in the interferometric data streams in the near 
future. The following plausible astrophysical considerations should provide motivations  for 
initiating such efforts. It was pointed out that such encounters involving neutron stars can give rise to 
certain resonant shattering flares in the electromagnetic sector 
due to strong tidal interactions between neutron stars during 
hyperbolic encounters,
though event rates are expected to be low \cite{DT13}.
Very recently, it was argued that aLIGO relevant  GW burst events may occur 
during hyperbolic encounters of Primordial Black Holes in dense clusters \cite{BN17}. 
Therefore, it should be of some interest to explore 
the search sensitivity and the possible false alarm rates of hyperbolic GW events by
adapting such an effort for eccentric inspirals \cite{Tiwari16}.

 The present computation will be crucial to obtain fully 3PN-accurate expressions for radiated energy 
 and angular momentum fluxes associated with hyperbolic encounters.
which are only available to 1PN-order \cite{BS89,JS92}.
Currently,  these computations are being extended to 3PN order \cite{BCG18}.
These investigation is expected to complement efforts that focus on 
the scatterings  of test particles by black hole space-time\cite{HC17,BG17,H17}.
It will also be desirable to adapt  Refs.~\cite{BD_2012, Damour16, BD17}
for exploring our GW burst signals using the framework of effective-one-body formalism.

\section*{Acknowledgments}
We thank Yannick Boetzel, Philippe Jetzer, Abhimanyu Susobhanan and Shubhanshu Tiwari for helpful discussions
and thank the anonymous reviewer for her/his  insightful comments and suggestions.

\newpage

\onecolumngrid
\appendix

\section{ Generalized quasi-Keplerian 
parametrization for hyperbolic compact binaries in ADM-type gauge }
\label{AppA}
We follow 
 exactly the same procedure, detailed in Sec.~\ref{secIIb}, while using 
 Eqs.~(A1) and (A2) of Ref.~\cite{MGS} 
to derive the 3PN accurate
hyperbolic parametrization in ADM-type gauge.
The third post-Newtonian accurate generalized quasi-Keplerian parametrization,
in ADM coordinates, for hyperbolic compact binaries 
is given by
\bes
\bea
r &=& a_r \left ( e_r\,\cosh u  -1 \right )\,,\\
\frac{2\pi}{P}(t-t_0) & =& e_t\sinh u -u+(\frac{f_{4t}}{c^4}+\frac{f_{6t}}{c^6})\, \nu+(\frac{g_{4t}}{c^4}+\frac{g_{6t}}{c^6})\sin \nu+\frac{h_{6t}}{c^6}\sin 2\,\nu+\frac{i_{6t}}{c^6}\sin 3\,\nu\,, \\
\frac{2\pi}{\Phi}(\phi-\phi_0) &=& \nu+(\frac{f_{4\phi}}{c^4}+\frac{f_{6\phi}}{c^6})\sin 2\,\nu+(\frac{g_{4\phi}}{c^4}+\frac{g_{6\phi}}{c^6})
\sin 3\,\nu+\frac{h_{6\phi}}{c^6}\sin 4\,\nu+\frac{i_{6\phi}}{c^6}\sin 5\,\nu\,,
\eea
\ens                       
where $ \nu  = 2\,\tanh^{-1} \biggl [ \biggl ( \frac{e_{\phi} +1 }{e_{\phi} -1 } \biggr )^{1/2} \, \tan \frac{u}{2} \biggr ]$.
The explicit 3PN accurate expressions for the orbital elements 
and functions of the
generalized quasi-Keplerian parametrization, in ADM coordinates,
read
\bes
\setlength{\jot}{10pt}
\bea
a_r &=&  \frac{1}{{(2\,E)}}\bigg\{ 1+\frac{ (2\, E )}{4\,c^2} \left( 7-\eta \right) +
\frac{{{ (2\, E) }}^{2}}{16 c^4}\,\bigg[ 
(1+10\,\eta+{\eta}^{2})
 \no&&
 +\frac {1}{(2\,E\,h^2)}
(68-44\,\eta)
\bigg] 
+{\frac {{{ (2\,E) }}^{3}}{192\,c^6}}\, 
\biggl [ 
-3+9\,\eta+6\,{\eta}^{2}
\no&&
-3\,{\eta}^{3}+\frac{1}{(2\,E\,h^2)}
\biggl (
864+ \left( -3\,{\pi}^{2}-2212 \right) \eta+432\,{\eta}^
{2}\biggr)
\no 
&&
+
\frac{1}
{ (2\,E\, h^2)^2 }
\biggl (
6432- \left( 13488-240\,{\pi}^{2} \right) \eta
+768\,{\eta}^{2}\biggr )
\biggr ]   \bigg\}\,,
\\
{e_{{r}}}^{2} &=&
1 + 2\,E\,h^2 +
\frac{(2\,E)}{4\,c^2}
\biggl \{ -24 +4
\,\eta+5\,\left(-3+ \eta \right) {(2\,E\,h^2)} 
\biggr \}
\no
&&
+ \frac{ (2\,E)^2}{8\,c^4}
\biggl \{
52+2\,\eta+2\,{\eta}^{2}
+\left( 80-55\,\eta+4\,{\eta}^{2} \right) {(2\,E\,h^2)}
\no&&
+\frac {8}{ (2\,E\,h^2)}
\left ( -17+11\, \eta \right )
\biggr \}
+ \frac{ (2\,E)^3 }{ 192\,c^6} \biggl \{
768+6\,\eta\,{\pi}^{2}
\no&&
+344\,\eta+216\,{\eta}^{2}
+ 3(2\,E\, h^2)\,\bigg(
-1488+1556\,\eta
-319\,{\eta}^{2}
\no&&
+4\,{\eta}^{3}
\bigg) \,
-\frac{4}{ (2\,E\, h^2)}\,\bigg(
588-8212\,\eta+177\,\eta
\,{\pi}^{2}+480\,{\eta}^{2}\bigg)
\no
&&
-\frac{192}{(2\,E\,h^2)^2}
\biggl (
134-281
\,\eta+5\,\eta\,{\pi}^{2}+16\,{\eta}^{2}
\biggr )
\biggr \}\,,
\\
n&=&{{(2\,E)}}^{3/2} \bigg\{ 1+{\frac {{(2\,E)}} {8\,{c}^{2}}}\, 
\left( -15+\eta \right)
+{\frac {{{(2\,E)}}^{2} }{128{c}^{4}}} 
\biggl [ 555 
+30\,\eta
+11\,{ \eta}^{2}
\biggr ]\no&&
+
{\frac {{{(2\,E)}}^{3}}{1024\,{c}^{6}}}
\biggl [  653+111\eta+7\eta^2+3\eta^3 
\biggr ] 
\bigg\} \,,
\\
{\it e_t}^2 &=&1+{2\,E}\,{h}^{2}+
{\frac {{(2\,E)}}{4\,{c}^{2}}}\, {\bigg\{8-8\,
\eta
+ \left( 17-7\,\eta \right) {(2\,E\, h^2)} \bigg\} }
\no&&
+
\frac{{{(2\,E)}}^{2}}{8\,{c}^{4}} \bigg\{8+4\,\eta +20\,{\eta}^{2}
+  {(2\,E\,h^2)}( 112-47\,\eta
+16\,{\eta}^{2} )
\no&&
+\frac{4}{(2\,E\,h^2)} \left( -17 + 11\,\eta \right )
\bigg\}
+{
\frac {{{(2\,E)}}^{3}}{192\,c^{6}}}
\bigg\{ 24\, \left(2-5\,\eta \right) 
 \left(-23+10\,\eta+ 4\,{\eta}^{2} \right) 
\no&&-15\,\biggl (-528
+200\,\eta-77\,{\eta}^{2}
+ 24\,{
\eta}^{3} \biggr ) {(2\,E\, h^2)}
\no&&- \frac{2}{(2\,E\,h^2)}
 \bigg(
6732 +117\,\eta\,{\pi }^{2}-12508\,\eta
+2004\,{\eta}^{2}\bigg)
\no&&
- \frac{96}{ (2\,E\,h^2)^2}
\bigg(
134 -281\,\eta+5\,\eta\,{\pi }^{2}+16\,{\eta}^{2}
\bigg)
 \bigg\}\,,
\\
f_{{4\,t}} &=&\frac{3\,(2\,E)^{2}}{2}\,\biggl \{ \frac{5 -2\,\eta }{ \sqrt{ (2\,E\,h^2)}}
 \biggr \}\,, 
\\
f_{{6\,t}} &=&{\frac {{{(2\,E)}}^{3}}{192(2Eh^2)^\frac{3}{2}}}
\biggl \{
\bigg(
10080+123\,\eta\,{\pi }^{2}-13952\,\eta
\no&&
+1440\,{\eta}^{2}
\bigg)
+(2Eh^2)36
\left (95-55\,\eta+18\,\eta^2
\right )
\biggr \}\,,
\\ 
g_{{4\,t}} &=& -\frac{1}{8}\,
\frac{ (2\,E)^2}{ \sqrt{ (2\,E\,h^2)}}
\biggl \{ (4 + \eta)\,\eta \, \sqrt{(1 +2\,E\,h^2)}
\biggr \}\,,
\\
g_{{6\,t}}&=&{\frac {{{(2\,E)}}^{3}}{192(2Eh^2)^\frac{3}{2}\sqrt{1+2Eh^2}}}
\bigg\{
3(2E h^2)^2 \eta \left(23 \eta^2-4 \eta-64\right)\no&&+(2E h^2)  \bigg(105 \eta^3+627 \eta^2+\left(3 \pi ^2-4232\right) \eta+1728\bigg)\no
&&+33 \eta^3+600 \eta^2+\left(3 \pi ^2-4148\right) \eta+1728
\bigg\}\,,
\\
h_{{6\,t}} &=&\frac{{{(2\,E)}}^{3}}{32}\,\eta\,
\biggl \{ 
\frac{(1 +2\,E\,h^2)}{ (2\,E\,h^2)^{3/2} }
\left(23+12\,\eta+ 6\,{\eta}^{2} \right) 
\biggr \}\,,
\\
i_{{6\,t}} &=&
{\frac {13\,{{(2\,E)}}^{3}}{192}}
\eta^3
\biggl ( 
\frac{  1 + 2\,E\,h^2 }{ 2\,E\,h^2}  
\biggr )^{3/2}\,,
\\
\Phi&=&2\,\pi \, \bigg\{ 1+{\frac {3}{{c}^{2}{h}^{2}}}+
-\frac{3(2E)^2}{4(2Eh^2)^2c^4}
\biggl [  -35+10\eta+(2Eh^2)\big(-5+2\eta\big)
\biggr ] 
\no&&
+{\frac {\,{{(2\,E)}}^{3}}{128\,{c}^{6}(2Eh^2)^3}}
\biggl [  
36960 +(615\pi^2-40000)\eta+1680\eta^2\no&&+  (2Eh^2)(10080+123\eta\pi^2-13952\eta+1440\eta^2)\no&&  + (2Eh^2)^2\big(120-120\eta+96\eta^2\big)
\biggr ] 
 \bigg\}\,, 
\\
f_{{4\,\phi}} &=&
\frac{{{(2\,E)}}^{2}}{8}
\,\frac{( 1+2\,E\,h^2)}{(2\,E\,h^2)^2}\,
\eta \, (1 -3\,\eta)\,,
\\
f_{{6\,\phi}}&=&{\frac {{{(2\,E)}}^{3}}{256(2Eh^2)^3}}
\bigg\{
256 + (-1076 + 49 \pi^2) \eta - 384 \eta^2 - 40 \eta^3 \no&&
+ 4 (2Eh^2)^2 \eta (-11 - 40 \eta + 24 \eta^2) + 
 (2Eh^2)\big(256 + (-1192 + 49 \pi^2) \eta\no&&- 336 \eta^2 + 80 \eta^3\big)
\bigg\}\,,
\\
g_{{4\,\phi}} &=&
-{\frac {3{{(2\,E)}}^{2}}{32}}
\frac{\,\eta^2\,}{(2\,E\,h^2)^2}
 ( 1 +2\,E\,h^2)^{3/2}\,,
\\
g_{{6\,\phi}}&=&
\frac{ (2\,E)^3}{768}\,\frac{ \sqrt{(1 +2\,E\,h^2)}}{(2Eh^2)^3}\,
\eta\bigg\{
220 + 3 \pi^2 + 96 \eta + 45 \eta^2 \no&&+ 3(2Eh^2)^2 \eta (-9 + 26 \eta) + 
 (2Eh^2)(220 + 3 \pi^2 + 312 \eta + 150 \eta^2)
\bigg\}\,,
\\
h_{{6\,\phi}} &=&{\frac {{{(2\,E)}}^{3}}{128}}
\,\frac{{(1 +2\,E\,h^2)}^2}{(2\,E\,h^2)^3}\,\eta 
\left( 5+28\,\eta+10\,{\eta}^{2} \right)\,,
\\
i_{{6\,\phi}} &=&
\frac{5\, (2\,E)^3}{256}\, \frac{\eta^3}{ (2\,E\,h^2)^3}
\, (1 +2\,E\,h^2)^{5/2} \,,
\\
{e_{{\phi}}}^{2}&=&
1 + 2\,E\,h^2 +
{\frac {{(2\,E)} }{4\,{c}^{2}}} \bigg\{ 
-24+ \left( -15+\eta \right) {(2\,E\,h^2)} \bigg\}
\no&&
+\frac{{{(2\,E)}}^{2} }{16\,{c}^{4}(2Eh^2)}
\bigg\{-408 +232\eta+15\eta^2+(2Eh^2)\big(-32+176\eta+18\eta^2\big)\no
&& +(2Eh^2)^2\big(160-30\eta+2\eta^2\big)\bigg\}
\no
&&
-{\frac {{{ (2\,E)}}^{3}}{384\,{c}^{6}(2Eh^2)^2}}
\bigg\{3 \big(27776 + (-65436 + 1325 \pi^2) \eta + 3440 \eta^2 - 70 \eta^3\big)\no&& + 
 36(2Eh^2)^3 (248 - 80 \eta + 13 \eta^2 + \eta^3) \no&&+ 
  6(2E h^2) (2456 + (-26860 + 581 \pi^2) \eta + 2689 \eta^2 + 10 \eta^3)\no&& + 
  (2Eh^2)^2 (-16032 + (2764 + 3 \pi^2) \eta + 4536 \eta^2 + 234 \eta^3)\biggr \}\,.
\eea
\ens

\onecolumngrid

\section{ Fully 3PN-accurate expressions for the dynamical variables that appear in the expressions for 
 $h_+|_Q(l) $ and $h_{\times}|_Q(l) $
 }
\label{AppB}


Extending the results we listed in Eq.~(\ref{eq:quasi_kepl_nonsp}), we provide 3PN-accurate expressions for 
$r, \dot r, \phi$ and $\dot \phi$ in terms of $\xi, e_t$ and $\eta$ in modified harmonic gauge.
The orbital separation reads 
\bes
\begin{align}
r\,=\,r_\text{N}+r_\text{1PN}+r_\text{2PN}+r_\text{3PN}\,,
\end{align}
where

\setlength{\jot}{9pt}
\bea
r_\text{N}\, &=&\, \frac{Gm}{c^2}\frac{1}{ \xi^{2/3}}(e_{\rm t}\cosh u-1)\,,\\
r_\text{1PN}\,&=&\,r_\text{N}\times\frac{\xi^{2/3} }{6 (e_t \cosh u-1)}[(7 \eta-6) e_t\cosh u+2 (\eta-9)]\,,\\
r_\text{2PN}\,&=&\,r_\text{N}\times\frac{\xi^{4/3} }{72 (e_t^2-1) (e_t \cosh u-1)}\big[(e_t^2-1) e_t (35 \eta^2-231 \eta+72) \cosh u-2 e_t^2 (4 \eta^2+15 \eta+36)+8 \eta^2+534 \eta\no&&-216\big]\,,\\
r_\text{3PN}\,&=&\,r_\text{N}\times\frac{\xi^2 }{181440 (e_t^2-1)^2 (e_t \cosh u-1)}\Big\{280 e_t^4 (16 \eta^3+90 \eta^2-81 \eta+432)+140 (e_t^2-1)^2 e_t (49 \eta^3-3933 \eta^2\no
&&\,+7047 \eta-864) \cosh u-e_t^2 \big[8960 \eta^3+3437280 \eta^2+81 (1435 \pi ^2-134336) \eta+3144960\big]+4480 \eta^3-761040 \eta^2\no
&&\,-348705 \pi ^2 \eta+12143736 \eta-4233600\Big\}\,.
\eea
\ens

The angular variable of the 3PN-accurate motion is given by
\bes
\begin{align}
\phi\,=\,\phi_\text{N}+\phi_\text{1PN}+\phi_\text{2PN}+\phi_\text{3PN}\,,
\end{align}
where

\setlength{\jot}{9pt}
\bea
\phi_\text{N}&=&\nu\,,\\
\phi_\text{1PN}&=&\frac{\xi^{2/3} }{(e_t^2-1) (e_t \cosh u-1)}\Big[e_t \sqrt{e_t^2-1} (4-\eta) \sinh u+3\eta(e_t\cosh u-1)  \Big]\,,\\
\phi_\text{2PN}&=&\frac{\xi^{4/3}}{192 (e_t^2-1)^{5/2} (e_t \cosh u-1)^2} \Bigg\{e_t (e_t^2-1) \Bigg[2 \Big\{e_t^2 \Big[384-\eta (7 \eta+275)\Big]+4 \Big[\eta (\eta+137)-792\Big]\Big\} \sinh u\\&&\,
+e_t \Big\{e_t^2 \Big[\eta (55 \eta-109)+384\Big]-4 \Big[\eta (13 \eta+41)-600\Big]\Big\} \sinh 2u\Bigg]+6 \sqrt{e_t^2-1} (e_t \cosh u-1)^2\noo\times
 \Big\{e_t^3 (1-3 \eta) \eta \sin 3\nu-8 \nu  \Big[e_t^2 (26 \eta-51)+28 \eta-78\Big]+4 e_t^2 \Big[(19-3 \eta) \eta+1\Big] \sin 2\nu\Big\}\Bigg\}\,,\no
\phi_\text{3PN}&=&\frac{\xi^2 }{53760 (e_t^2-1)^{7/2} (e_t \cosh u-1)^3}\Bigg(\sqrt{e_t^2-1} (e_t \cosh u-1)^3 (2 e_t^2 \Bigg\{\Bigg[280 e_t^2 \Big\{\eta \Big[\eta (93 \eta-781)+886\Big]+24\Big\}\\&&\,
+\eta \Big\{32 \Big[35 \eta (9 \eta-395)+36877\Big]-30135 \pi ^2\Big\}+84000\Bigg] \sin 2\nu+e_t \eta \Bigg[\Big\{35 e_t^2 \Big[\eta (129 \eta-137)+33\Big]+4 \Big[35\noo\times
 \eta (51 \eta-727)+28302\Big]-4305 \pi ^2\Big\} \sin 3\nu+35 e_t \Big\{3 e_t \Big[5 (\eta-1) \eta+1\Big] \sin 5\nu+4 \Big[3 \eta (5 \eta-19)+82\Big] \sin 4\nu\noo
\Big\}\Bigg]\Bigg\}+420 \nu  \Big\{16 (65 e_t^4+320 e_t^2+56) \eta^2+\Big[123 \pi ^2 (e_t^2+4)-32 (55 e_t^4+870 e_t^2+793)\Big] \eta+96 (26 e_t^4+293 e_t^2
\noo
+190)\Big\})+e_t (e_t^2-1) \sinh u\Big\{-70 e_t^6 \eta \Big[\eta (71 \eta+61)-639\Big]+1680 e_t^4 (\eta-4) \cosh ^2u \Big\{3 e_t \eta (3 \eta-1) \cos 3\nu\noo
+8 \Big[\eta (3 \eta-19)-1\Big] \cos 2\nu\Big\}+e_t^4 \Bigg[\eta \Big\{4 \Big[70 \eta (125 \eta-507)-462853\Big]-4305 \pi ^2\Big\}+3933440\Bigg]+1680 e_t^2 (\eta-4)
\noo
\times \Big\{3 e_t \eta (3 \eta-1) \cos 3\nu+8 \Big[\eta (3 \eta-19)-1\Big] \cos 2\nu\Big\}+e_t^2 \Big\{6 \eta \Big[140 \eta (25 \eta-397)+1435 \pi ^2-917424\Big]\noo+7947520\Big\}+4 e_t \cosh u \Bigg\{-70 e_t^4 \Big\{\eta \Big[\eta (39 \eta-719)+2279\Big]-3072\Big\}+840 e_t^2 (\eta-4) \Big\{3 e_t (1-3 \eta) \eta \cos 3\nu\noo+8 \Big[(19-3 \eta) \eta
+1\Big] \cos 2\nu\Big\}+e_t^2 \Bigg[\eta \Big\{8 \Big[35 (232-53 \eta) \eta+186959\Big]+4305 \pi ^2\Big\}-2983680\Bigg]-20 \Bigg[323904\noo
-\eta \Big\{4 \Big[7 \eta (\eta+45)+56013\Big]-861 \pi ^2\Big\}\Bigg]\Bigg\}+e_t^2 \Bigg\{-70 e_t^4 \eta \Big[\eta (71 \eta+61)-639\Big]+e_t^2 \Bigg[\eta \Big\{4 \Big[35 \eta (229 \eta-1173)\noo-384978\Big]-4305 \pi ^2\Big\}+3646720\Bigg]+20 \Bigg[\eta \Big\{861 \pi ^2-4 \Big[14 \eta (9 \eta-25)+54025\Big]\Big\}+280000\Bigg]\Bigg\} \cosh 2u\noo+40 \Big[\eta (1456 \eta+861 \pi ^2
-253508)+396480\Big]\Big\}\Bigg)\,,\no
\eea
\ens
with $\nu = 2\arctan(\sqrt{\frac{e_t^2+1}{e_t^2-1}}\tanh\frac{u}{2})$ as before. Furthermore, we require explicit expressions for the first time derivatives 
$\dot r$ and $\dot \phi$ to compute GW waveforms from binaries in hyperbolic orbits, namely,
\bes
\setlength{\jot}{9pt}
\bea
\dot r\,&=&\,\dot r_\text{N}+\dot r_\text{1PN}+\dot r_\text{2PN}+\dot r_\text{3PN}\,,\\
\dot r_\text{N}&=&{\xi}^{1/3}\frac{c\, e_t\, \sinh u}{e_t\cosh u-1}\,,\\
\dot r_\text{1PN}&=&\dot r_\text{N}\times\frac{\xi^{2/3}}{6} \ (7 \eta-6)\,,\\
\dot r_\text{2PN}&=&\dot r_\text{N}\times\frac{\xi^{4/3} }{72 (e_t \cosh u-1)^2}\Big\{9 e_t (\eta-15) \eta \cos \nu +e_t \Big[7 \eta (5 \eta-33)+72\Big] \cosh u (e_t \cosh u-2)+5 \eta (7 \eta-3)\noo-468\Big\}\,,\\
\dot r_\text{3PN}&=&\dot r_\text{N}\times\frac{\xi^2 }{181440 (e_t^2-1)^{3/2} (e_t \cosh u-1)^3}\Bigg\{3780 (e_t^2-1)^{3/2} (7 \eta-6) (e_t \cosh u-1) \Big[e_t (\eta-15) \eta \cos \nu+24 \eta\noo
-60\Big]+140 (e_t^2-1)^{3/2} (49 \eta^3-3933 \eta^2+7047 \eta-864) (e_t \cosh u-1)^3-27 \Bigg[-840 (e_t^2-1) e_t^2 \eta (\eta^2-19 \eta\noo
+60) \sin \nu \sinh u-840 \sqrt{e_t^2-1} e_t^3 \eta (\eta^2-19 \eta+60) \cos \nu \sinh ^2u+840 \sqrt{e_t^2-1} e_t^3 \eta (\eta^2-19 \eta+60) \cos \nu \noo
\times\cosh ^2u+\sqrt{e_t^2-1} e_t \cosh u \Big\{e_t \Big[35 (65 e_t^2-32) \eta^3-525 (27 e_t^2+88) \eta^2+(-315 e_t^2-4305 \pi ^2+93468) \eta\noo
+67200\Big] \cos \nu+35 \Big[3 e_t^3 \eta (13 \eta^2-73 \eta+23) \cos 3\nu+24 e_t^2 \eta (3 \eta^2-49 \eta+116) \cos 2\nu+1056 e_t^2 \eta^2-2784 e_t^2 \eta\noo
+2880 e_t^2+384 \eta^2+123 \pi ^2 \eta-9440 \eta+2880\Big]\Big\}-\sqrt{e_t^2-1} \Big\{e_t \Big[35 (65 e_t^2-8) \eta^3-105 (135 e_t^2+592) \eta^2-3 \noo
\times(105 e_t^2+1435 \pi ^2-47956) \eta+67200\Big] \cos \nu+35 \Big[3 e_t^3 \eta (13 \eta^2-73 \eta+23) \cos 3\nu+24 e_t^2 \eta (3 \eta^2-49 \eta+116) \noo
\times\cos 2\nu+480 e_t^2 \eta^2+960 e_t^2 \eta-2880 e_t^2+960 \eta^2+123 \pi ^2 \eta-13184 \eta+8640\Big]\Big\}\Bigg]\Bigg\}\,,
\eea
\ens

as well as
\bes
\setlength{\jot}{9pt}
\bea
\dot \phi\,&=&\,\dot\phi_\text{N}+\dot\phi_\text{1PN}+\dot\phi_\text{2PN}+\dot\phi_\text{3PN}\,,\\
\dot \phi_\text{N}&=&\frac{n \sqrt{e_t^2-1}}{(e_t \cosh u-1)^2}\,,\\
\dot \phi_\text{1PN}&=&\dot\phi_\text{N}\times\frac{\xi^{2/3}}{(e_t^2-1) (e_t \cosh u-1)}\Big[e_t^2 (\eta-4)+e_t (\eta-1) \cosh u-3\Big]\,,\\
\dot \phi_\text{2PN}&=&\dot\phi_\text{N}\times\frac{\xi^{4/3} }{192 (e_t^2-1)^2 (e_t \cosh u-1)^2}\Big\{-6 e_t^4 \cosh ^2u \Big[3 e_t \eta (3 \eta-1) \cos 3\nu+8 (3 \eta^2-19 \eta-1) \cos 2\nu\Big]-e_t^2\, \Big[\,\noo
 \times e_t^2 \,(103 \eta^2+131 \eta-72)-4 (25 \eta^2-223 \eta+60)\Big] \cosh 2u+2 e_t \cosh u \Big[55 e_t^4 \eta^2-109 e_t^4 \eta+384 e_t^4+18 e_t^3 \eta \noo
\times(3 \eta-1) \cos 3\nu+48 e_t^2 (3 \eta^2-19 \eta-1) \cos 2\nu-45 e_t^2 \eta^2+1359 e_t^2 \eta-432 e_t^2-4 \eta^2+796 \eta-576\Big]+3 \,\Big[\noo
-7 e_t^4 \eta^2-291 e_t^4 \eta+312 e_t^4-18 e_t^3 \eta^2 \cos 3\nu+6 e_t^3 \eta \cos 3\nu+16 e_t^2 (-3 \eta^2+19 \eta+1) \cos 2\nu+8 (e_t^2-1)^2 e_t (\eta\noo
-15) \eta \cos \nu+4 e_t^2 \eta^2-476 e_t^2 \eta-768 e_t^2-256 \eta+768\Big]\Big\}\,,\\
\dot \phi_\text{3PN}&=&\dot\phi_\text{N}\times\frac{\xi^2 }{107520 (e_t^2-1)^3 (e_t \cosh u-1)^3}\Bigg(140 \eta \Big\{16 \Big[3 \eta (5 \eta-19)+82\Big] \cos 4\nu+15 e_t \Big[5 (\eta-1) \eta+1\Big] \cos 5\nu\Big\}\noo
\times\cosh ^3u\, e_t^7+4 \Bigg\{3 e_t \eta \Bigg[-525 \Big[5 (\eta-1) \eta+1\Big] \cos 5\nu e_t^2-560 \Big[3 \eta  (5 \eta-19)+82\Big] \cos 4\nu e_t+2 \Big\{-35 \Big[\eta \noo
\times(93 \eta+19)-15\Big] e_t^2-8 \Big[35 \eta (21 \eta-344)+13941\Big]+4305 \pi ^2\Big\} \cos 3\nu\Bigg]-4 \Bigg[\eta \Big\{280 (75 \eta^2-595 \eta+436) e_t^2\noo
+16 \Big[35 \eta (9 \eta-697)+65879\Big]-30135 \pi ^2\Big\}+77280\Bigg] \cos 2\nu\Bigg\} \cosh ^2u e_t^4+\Bigg[70 \Big\{\eta \Big[\eta (291 \eta+1865)-2639\Big]\noo
+1344\Big\} e_t^4+\Big\{\eta \Big[140 (10271-475 \eta) \eta+30135 \pi ^2-4218008\Big]+934080\Big\} e_t^2-240 (5521 \eta+56)+140 \eta \Big[16\noo
\times \eta\, (13 \eta+24)+615 \pi ^2\Big]\Bigg] \cosh 3u e_t^3+2 \Bigg\{-3 \eta \Big\{35 \Big[\eta (129 \eta-137)+33\Big] e_t^2+1540 \eta (3 \eta-59)-4305 \pi ^2\noo
+109848\Big\} \cos 3\nu\, e_t^3-2 \Bigg[280 \Big\{\eta \Big[\eta (93 \eta-781)+886\Big]+24\Big\} e_t^2+(-338240 \eta-30135 \pi ^2+928064) \eta+70560\Bigg]\noo
\times\cos 2\nu\, e_t^2-70 (71 e_t^6-572 e_t^4-260 e_t^2+32) \eta^3-70 (61 e_t^6+14900 e_t^4+44228 e_t^2+10336) \eta^2+6720 (138 e_t^4\noo
-377 e_t^2-214)+\Big[44730 e_t^6+518508 e_t^4+8417056 e_t^2-4305 (e_t^2+8)^2 \pi ^2+8203040\Big] \eta\Bigg\} \cosh 2u\, e_t^2+\cosh u \noo
\times \bigg(-70 \,\Big\{\eta\, \Big[\eta (319 \eta-8163)+16997\Big]-30144\Big\} e_t^6+\Bigg[\eta \Big\{55965 \pi ^2-4 \Big[35 \eta (1331 \eta-56471)+953852\Big]\Big\}\noo
-7627200\Bigg] e_t^4+2 \Bigg\{2 \Bigg[280 \Big\{\eta \Big[\eta (93 \eta-781)+886\Big]+24\Big\} e_t^2+(-338240 \eta-30135 \pi ^2+928064) \eta+70560\Bigg] \noo
\times\cos 2\nu+3 e_t \eta \Big\{35 \Big[\eta (129 \eta-137)+33\Big] e_t^2+1540 \eta (3 \eta-59)-4305 \pi ^2+109848\Big\} \cos 3\nu\Bigg\} \cosh 2u\, e_t^4\noo
+56 \Big\{\eta \Big[10 (24299-83 \eta) \eta+10455 \pi ^2-739322\Big]+130320\Big\} e_t^2+2\Bigg (e_t \bigg(2 \Bigg\{280 \Big\{\eta \Big[\eta (57 \eta-193)-506\Big]+24\Big\} \noo
\times e_t^4+\Bigg[\eta \Big\{16 \Big[35 \eta (243 \eta-2791)+136754\Big]-30135 \pi ^2\Big\}+57120\Bigg] e_t^2+6 \Bigg[\eta \Big\{16 \Big[35 \eta (9 \eta-679)+64444\Big]\noo
-30135 \pi ^2\Big\}+79520\Bigg]\Bigg\} \cos 2\nu+3 e_t \eta \Bigg\{\Bigg[35 \Big[\eta (25 \eta+447)-151\Big] e_t^4+\Big\{2 \Big[35 \eta (413 \eta-1669)+58109\Big]\noo
-4305 \pi ^2\Big\} e_t^2+8 \Big[70 \eta (61 \eta-1015)+83261\Big]-25830 \pi ^2\Bigg] \cos 3\nu+70 e_t \Big\{16 \Big[3 \eta (5 \eta-19)+82\Big] \cos 4\nu+15 e_t \noo
\times\Big[5 (\eta-1) \eta+1\Big] \cos 5\nu\Big\}\Bigg\}\bigg)-8 (e_t^2-1)^2 \Bigg[\eta \Big\{35 \Big[5 \eta (13 \eta-81)-9\Big] e_t^2-4 \Big[70 \eta (7 \eta+117)-20217\Big]\noo-4305 \pi ^2\Big\}+67200\Bigg] \cos \nu\Bigg) e_t+320 \Big[\eta (7028 \eta+3444 \pi ^2-123467)+42000\Big]\bigg) e_t+2 \bigg(-70 (71 e_t^4-542 e_t^2\noo
-744) \eta^3\, e_t^4
+\Bigg(8 \Bigg[\eta \Big\{35 \Big[\eta (17 \eta+507)-2889\Big] e_t^2-4 \Big[70 \eta (\eta+231)-45417\Big]-4305 \pi ^2\Big\}+67200\Bigg] \cos \nu\noo
\times (e_t^2-1)^2+e_t (-2 \Bigg\{280 \Big\{\eta \Big[\eta (57 \eta-193)-506\Big]+24\Big\} e_t^4+\Bigg[\eta \Big\{16 \Big[35 \eta (93 \eta-1601)+106234\Big]-30135 \pi ^2\Big\}\noo
+57120\Bigg] e_t^2+64 (30787 \eta+2625)+70 \eta \Big[16 \eta (9 \eta-643)-861 \pi ^2\Big]\Bigg\} \cos 2\nu+e_t \eta \Bigg\{3 \Bigg[-35 \Big[\eta (25 \eta+447)-151\Big]\noo
\times e_t^4+\Big\{4305 \pi ^2-2 \Big[35 \eta (227 \eta-1707)+59159\Big]\Big\} e_t^2+2 \Big[280 (327-19 \eta) \eta+4305 \pi ^2-109988\Big]\Bigg] \cos 3\nu+70 e_t \noo
\times\Big\{-16 \Big[3 \eta (5 \eta-19)+82\Big] \cos 4\nu-15 e_t \Big[5 (\eta-1) \eta+1\Big] \cos 5\nu\Big\}\Bigg\}-840 \sqrt{e_t^2-1} (\eta-4) \Big\{64 e_t \Big[\eta (3 \eta-19)\noo
-1\Big] \cos \nu (e_t \cosh u-1)^2+36 e_t^2 \eta (3 \eta-1) \cos 2\nu (e_t \cosh u-1)^2+\eta \Big[(19 \eta+111) e_t^4+9 (3 \eta-1) (e_t \cosh 2u\noo
-4 \cosh u) e_t^3+(70 \eta-258) e_t^2-8 (\eta-15)\Big]\Big\} \sin \nu \sinh u)\Bigg) e_t-70 (61 e_t^8+11454 e_t^6+57640 e_t^4+45184 e_t^2\noo
+1536) \eta^2+6720 (34 e_t^6-e_t^4-188 e_t^2-600)+\Big[44730 e_t^8+263008 e_t^6+4781712 e_t^4+17066880 e_t^2-4305 (e_t^6\noo
+10 e_t^4+84 e_t^2+40) \pi ^2+6482560\Big] \eta\bigg)\Bigg)\,.\no
\eea
\ens
\section{Relations between coefficients in the parametrization of $t-t_0$ and $\phi-\phi_0$}
\label{AppC}
Our 3PN-accurate Keplerian-type parametric solution, derived from first principles, relies on explicit expressions for certain coefficients $c_i, c_i'$ and $e_i, e_i'$ 
to parametrize the radial and angular motion, respectively. In Sec.~II.B, we have used the following explicit relations between coefficients $c_i$ and 
$c'_i = b_i/( a_r^{i-1}\, \sqrt{-s_+\, s_-})$ to obtain the parametric solution for $t-t_0$ in Eq.~(\ref{temp_t}) from Eq.~(\ref{tt0_ciprime}).
\bes
\begin{align}
c_0= &\,c'_0\,e_r\,,\\
c_1=& \,c'_0 -c'_1\,,\\
c_2= &\,\frac{c_2'}{(e_r^2-1)^{1/2}}+ \frac{c_3'}{(e_r^2-1)^{3/2}}+\frac{c'_4\, (e_r^2+2)}{2\,(e_r^2-1)^{5/2}}+\frac{c'_5}{(e_r^2-1)^{7/2}}(1+\frac{3e_r^2}{2})\,,\\
c_3=&\,\frac{c_3'\,e_r}{(e_r^2-1)^{3/2}}+\frac{2\,c'_4\,e_r}{\,(e_r^2-1)^{5/2}}+\frac{c_5'}{(e_r^2-1)^{7/2}}\,(3\,e_r+\frac{3}{4}\,e_r^3)\,,\\
c_4=&\,\frac{\,c'_4\,e_r^2}{4\,(e_r^2-1)^{5/2}}+\frac{3\,c_5'\,e_r^2}{4\,(e_r^2-1)^{7/2}}\,,\\
c_5=&\,\frac{\,c_5'\,e_r^3}{12\,(e_r^2-1)^{7/2}}\,.
\end{align}
\ens

Also in Sec.~II.B, the parametric solution for $\phi-\phi_0$ in Eq.~(\ref{Eq.2.29}) was obtained from Eq.~(\ref{phiphi0_eiprime}) by using explicit relations between the coefficients $e_i$ and 
$e'_i = d_i/( a_r^{i+1}\,\sqrt{- s_+s_-})$, namely:
\bes
\begin{align}
e_0= &\frac{{e_0'}}{({e_r}^2-1)^{1/2}}+\frac{{e_1'}}{\left({e_r}^2-1\right)^{3/2}}+\frac{{e_2'} \left({e_r}^2+2\right)}{2 \left({e_r}^2-1\right)^{5/2}}+\frac{ {e_3'}\,(3 {e_r}^2+2)}{2 \left({e_r}^2-1\right)^{7/2}}+\frac{{e_4'} \left(3 {e_r}^4+24 {e_r}^2+8\right)}{8 \left({e_r}^2-1\right)^{9/2}}+\frac{{e_5'} \left(15 {e_r}^4+40 {e_r}^2+8\right)}{8 \left({e_r}^2-1\right)^{11/2}},\\
e_1=&\frac{{e_1'} {e_r}}{\left({e_r}^2-1\right)^{3/2}}+\frac{2 {e_2'} {e_r}}{\left({e_r}^2-1\right)^{5/2}}+\frac{3 {e_3'} \left({e_r}^2+4\right) {e_r}}{4 \left({e_r}^2-1\right)^{7/2}}+\frac{{e_4'} \left(3 {e_r}^2+4\right) {e_r}}{\left({e_r}^2-1\right)^{9/2}}+\frac{5 {e_5'} \left({e_r}^4+12 {e_r}^2+8\right) {e_r}}{8 \left({e_r}^2-1\right)^{11/2}}\,,\\
e_2= &\frac{{e_2'} {e_r}^2}{4 \left({e_r}^2-1\right)^{5/2}}+\frac{3 {e_3'} {e_r}^2}{4 \left({e_r}^2-1\right)^{7/2}}+\frac{{e_4'} \left({e_r}^2+6\right) {e_r}^2}{4 \left({e_r}^2-1\right)^{9/2}}+\frac{5 {e_5'} \left({e_r}^2+2\right) {e_r}^2}{4 \left({e_r}^2-1\right)^{11/2}}\,,\\
e_3=&\frac{{e_3'} {e_r}^3}{12 \left({e_r}^2-1\right)^{7/2}}+\frac{{e_4'} {e_r}^3}{3 \left({e_r}^2-1\right)^{9/2}}+\frac{5 {e_5'} \left({e_r}^2+8\right) {e_r}^3}{48 \left({e_r}^2-1\right)^{11/2}}\,,\\
e_4=&\frac{{e_4'} {e_r}^4}{32 \left({e_r}^2-1\right)^{9/2}}+\frac{5 {e_5'} {e_r}^4}{32 \left({e_r}^2-1\right)^{11/2}}\,,\\
e_5=&\frac{{e_5'} {e_r}^5}{80 \left({e_r}^2-1\right)^{11/2}}\,.
\end{align}
\ens
\newpage

\bibliography{GGL}

\begin{thebibliography}{46}
\expandafter\ifx\csname natexlab\endcsname\relax\def\natexlab#1{#1}\fi
\expandafter\ifx\csname bibnamefont\endcsname\relax
  \def\bibnamefont#1{#1}\fi
\expandafter\ifx\csname bibfnamefont\endcsname\relax
  \def\bibfnamefont#1{#1}\fi
\expandafter\ifx\csname citenamefont\endcsname\relax
  \def\citenamefont#1{#1}\fi
\expandafter\ifx\csname url\endcsname\relax
  \def\url#1{\texttt{#1}}\fi
\expandafter\ifx\csname urlprefix\endcsname\relax\def\urlprefix{URL }\fi
\providecommand{\bibinfo}[2]{#2}
\providecommand{\eprint}[2][]{\url{#2}}
\newcommand{\apjl}{The Astrophysical Journal Letters}
\newcommand{\mnras}{Monthly Notices of the Royal Astronomical Society}

\bibitem[{\citenamefont{{Abbott}
  et~al.}(2016{\natexlab{a}})\citenamefont{{Abbott}, {Abbott}, {Abbott},
  {Abernathy}, {Acernese}, {Ackley}, {Adams}, {Adams}, {Addesso}, {Adhikari}
  et~al.}}]{aLIGO}
\bibinfo{author}{\bibfnamefont{B.~P.} \bibnamefont{{Abbott}}},
  \bibinfo{author}{\bibfnamefont{R.}~\bibnamefont{{Abbott}}},
  \bibinfo{author}{\bibfnamefont{T.~D.} \bibnamefont{{Abbott}}},
  \bibinfo{author}{\bibfnamefont{M.~R.} \bibnamefont{{Abernathy}}},
  \bibinfo{author}{\bibfnamefont{F.}~\bibnamefont{{Acernese}}},
  \bibinfo{author}{\bibfnamefont{K.}~\bibnamefont{{Ackley}}},
  \bibinfo{author}{\bibfnamefont{C.}~\bibnamefont{{Adams}}},
  \bibinfo{author}{\bibfnamefont{T.}~\bibnamefont{{Adams}}},
  \bibinfo{author}{\bibfnamefont{P.}~\bibnamefont{{Addesso}}},
  \bibinfo{author}{\bibfnamefont{R.~X.} \bibnamefont{{Adhikari}}},
  \bibnamefont{et~al.}, \bibinfo{journal}{Physical Review Letters}
  \textbf{\bibinfo{volume}{116}}, \bibinfo{eid}{131103}
  (\bibinfo{year}{2016}{\natexlab{a}}), \eprint{1602.03838}.

\bibitem[{\citenamefont{{Abbott}
  et~al.}(2016{\natexlab{b}})\citenamefont{{Abbott}, {Abbott}, {Abbott},
  {Abernathy}, {Acernese}, {Ackley}, {Adams}, {Adams}, {Addesso}, {Adhikari}
  et~al.}}]{GW_1_d}
\bibinfo{author}{\bibfnamefont{B.~P.} \bibnamefont{{Abbott}}},
  \bibinfo{author}{\bibfnamefont{R.}~\bibnamefont{{Abbott}}},
  \bibinfo{author}{\bibfnamefont{T.~D.} \bibnamefont{{Abbott}}},
  \bibinfo{author}{\bibfnamefont{M.~R.} \bibnamefont{{Abernathy}}},
  \bibinfo{author}{\bibfnamefont{F.}~\bibnamefont{{Acernese}}},
  \bibinfo{author}{\bibfnamefont{K.}~\bibnamefont{{Ackley}}},
  \bibinfo{author}{\bibfnamefont{C.}~\bibnamefont{{Adams}}},
  \bibinfo{author}{\bibfnamefont{T.}~\bibnamefont{{Adams}}},
  \bibinfo{author}{\bibfnamefont{P.}~\bibnamefont{{Addesso}}},
  \bibinfo{author}{\bibfnamefont{R.~X.} \bibnamefont{{Adhikari}}},
  \bibnamefont{et~al.}, \bibinfo{journal}{Physical Review Letters}
  \textbf{\bibinfo{volume}{116}}, \bibinfo{eid}{061102}
  (\bibinfo{year}{2016}{\natexlab{b}}), \eprint{1602.03837}.

\bibitem[{\citenamefont{{Abbott}
  et~al.}(2016{\natexlab{c}})\citenamefont{{Abbott}, {Abbott}, {Abbott},
  {Abernathy}, {Acernese}, {Ackley}, {Adams}, {Adams}, {Addesso}, {Adhikari}
  et~al.}}]{GW_2_d}
\bibinfo{author}{\bibfnamefont{B.~P.} \bibnamefont{{Abbott}}},
  \bibinfo{author}{\bibfnamefont{R.}~\bibnamefont{{Abbott}}},
  \bibinfo{author}{\bibfnamefont{T.~D.} \bibnamefont{{Abbott}}},
  \bibinfo{author}{\bibfnamefont{M.~R.} \bibnamefont{{Abernathy}}},
  \bibinfo{author}{\bibfnamefont{F.}~\bibnamefont{{Acernese}}},
  \bibinfo{author}{\bibfnamefont{K.}~\bibnamefont{{Ackley}}},
  \bibinfo{author}{\bibfnamefont{C.}~\bibnamefont{{Adams}}},
  \bibinfo{author}{\bibfnamefont{T.}~\bibnamefont{{Adams}}},
  \bibinfo{author}{\bibfnamefont{P.}~\bibnamefont{{Addesso}}},
  \bibinfo{author}{\bibfnamefont{R.~X.} \bibnamefont{{Adhikari}}},
  \bibnamefont{et~al.}, \bibinfo{journal}{Physical Review Letters}
  \textbf{\bibinfo{volume}{116}}, \bibinfo{eid}{241103}
  (\bibinfo{year}{2016}{\natexlab{c}}), \eprint{1606.04855}.

\bibitem[{\citenamefont{{Abbott}
  et~al.}(2017{\natexlab{a}})\citenamefont{{Abbott}, {Abbott}, {Abbott},
  {Acernese}, {Ackley}, {Adams}, {Adams}, {Addesso}, {Adhikari}, {Adya}
  et~al.}}]{GW_3_d}
\bibinfo{author}{\bibfnamefont{B.~P.} \bibnamefont{{Abbott}}},
  \bibinfo{author}{\bibfnamefont{R.}~\bibnamefont{{Abbott}}},
  \bibinfo{author}{\bibfnamefont{T.~D.} \bibnamefont{{Abbott}}},
  \bibinfo{author}{\bibfnamefont{F.}~\bibnamefont{{Acernese}}},
  \bibinfo{author}{\bibfnamefont{K.}~\bibnamefont{{Ackley}}},
  \bibinfo{author}{\bibfnamefont{C.}~\bibnamefont{{Adams}}},
  \bibinfo{author}{\bibfnamefont{T.}~\bibnamefont{{Adams}}},
  \bibinfo{author}{\bibfnamefont{P.}~\bibnamefont{{Addesso}}},
  \bibinfo{author}{\bibfnamefont{R.~X.} \bibnamefont{{Adhikari}}},
  \bibinfo{author}{\bibfnamefont{V.~B.} \bibnamefont{{Adya}}},
  \bibnamefont{et~al.}, \bibinfo{journal}{Physical Review Letters}
  \textbf{\bibinfo{volume}{118}}, \bibinfo{eid}{221101}
  (\bibinfo{year}{2017}{\natexlab{a}}), \eprint{1706.01812}.

\bibitem[{\citenamefont{{Abbott}
  et~al.}(2017{\natexlab{b}})\citenamefont{{Abbott}, {Abbott}, {Abbott},
  {Acernese}, {Ackley}, {Adams}, {Adams}, {Addesso}, {Adhikari}, {Adya}
  et~al.}}]{GW_6_d}
\bibinfo{author}{\bibfnamefont{B.~P.} \bibnamefont{{Abbott}}},
  \bibinfo{author}{\bibfnamefont{R.}~\bibnamefont{{Abbott}}},
  \bibinfo{author}{\bibfnamefont{T.~D.} \bibnamefont{{Abbott}}},
  \bibinfo{author}{\bibfnamefont{F.}~\bibnamefont{{Acernese}}},
  \bibinfo{author}{\bibfnamefont{K.}~\bibnamefont{{Ackley}}},
  \bibinfo{author}{\bibfnamefont{C.}~\bibnamefont{{Adams}}},
  \bibinfo{author}{\bibfnamefont{T.}~\bibnamefont{{Adams}}},
  \bibinfo{author}{\bibfnamefont{P.}~\bibnamefont{{Addesso}}},
  \bibinfo{author}{\bibfnamefont{R.~X.} \bibnamefont{{Adhikari}}},
  \bibinfo{author}{\bibfnamefont{V.~B.} \bibnamefont{{Adya}}},
  \bibnamefont{et~al.}, \bibinfo{journal}{\apjl}
  \textbf{\bibinfo{volume}{851}}, \bibinfo{eid}{L35}
  (\bibinfo{year}{2017}{\natexlab{b}}).

\bibitem[{\citenamefont{{Abbott}
  et~al.}(2017{\natexlab{c}})\citenamefont{{Abbott}, {Abbott}, {Abbott},
  {Acernese}, {Ackley}, {Adams}, {Adams}, {Addesso}, {Adhikari}, {Adya}
  et~al.}}]{GW_4_d}
\bibinfo{author}{\bibfnamefont{B.~P.} \bibnamefont{{Abbott}}},
  \bibinfo{author}{\bibfnamefont{R.}~\bibnamefont{{Abbott}}},
  \bibinfo{author}{\bibfnamefont{T.~D.} \bibnamefont{{Abbott}}},
  \bibinfo{author}{\bibfnamefont{F.}~\bibnamefont{{Acernese}}},
  \bibinfo{author}{\bibfnamefont{K.}~\bibnamefont{{Ackley}}},
  \bibinfo{author}{\bibfnamefont{C.}~\bibnamefont{{Adams}}},
  \bibinfo{author}{\bibfnamefont{T.}~\bibnamefont{{Adams}}},
  \bibinfo{author}{\bibfnamefont{P.}~\bibnamefont{{Addesso}}},
  \bibinfo{author}{\bibfnamefont{R.~X.} \bibnamefont{{Adhikari}}},
  \bibinfo{author}{\bibfnamefont{V.~B.} \bibnamefont{{Adya}}},
  \bibnamefont{et~al.}, \bibinfo{journal}{Physical Review Letters}
  \textbf{\bibinfo{volume}{119}}, \bibinfo{eid}{141101}
  (\bibinfo{year}{2017}{\natexlab{c}}), \eprint{1709.09660}.

\bibitem[{\citenamefont{{Abbott}
  et~al.}(2017{\natexlab{d}})\citenamefont{{Abbott}, {Abbott}, {Abbott},
  {Acernese}, {Ackley}, {Adams}, {Adams}, {Addesso}, {Adhikari}, {Adya}
  et~al.}}]{GW_5_d}
\bibinfo{author}{\bibfnamefont{B.~P.} \bibnamefont{{Abbott}}},
  \bibinfo{author}{\bibfnamefont{R.}~\bibnamefont{{Abbott}}},
  \bibinfo{author}{\bibfnamefont{T.~D.} \bibnamefont{{Abbott}}},
  \bibinfo{author}{\bibfnamefont{F.}~\bibnamefont{{Acernese}}},
  \bibinfo{author}{\bibfnamefont{K.}~\bibnamefont{{Ackley}}},
  \bibinfo{author}{\bibfnamefont{C.}~\bibnamefont{{Adams}}},
  \bibinfo{author}{\bibfnamefont{T.}~\bibnamefont{{Adams}}},
  \bibinfo{author}{\bibfnamefont{P.}~\bibnamefont{{Addesso}}},
  \bibinfo{author}{\bibfnamefont{R.~X.} \bibnamefont{{Adhikari}}},
  \bibinfo{author}{\bibfnamefont{V.~B.} \bibnamefont{{Adya}}},
  \bibnamefont{et~al.}, \bibinfo{journal}{Physical Review Letters}
  \textbf{\bibinfo{volume}{119}}, \bibinfo{eid}{161101}
  (\bibinfo{year}{2017}{\natexlab{d}}), \eprint{1710.05832}.

\bibitem[{\citenamefont{{Taylor}}(1993)}]{JT_1993}
\bibinfo{author}{\bibfnamefont{J.~H.} \bibnamefont{{Taylor}}},
  \bibinfo{journal}{Classical and Quantum Gravity}
  \textbf{\bibinfo{volume}{10}}, \bibinfo{pages}{S167} (\bibinfo{year}{1993}).

\bibitem[{\citenamefont{{Lyne} et~al.}(2004)\citenamefont{{Lyne}, {Burgay},
  {Kramer}, {Possenti}, {Manchester}, {Camilo}, {McLaughlin}, {Lorimer},
  {D'Amico}, {Joshi} et~al.}}]{J0737}
\bibinfo{author}{\bibfnamefont{A.~G.} \bibnamefont{{Lyne}}},
  \bibinfo{author}{\bibfnamefont{M.}~\bibnamefont{{Burgay}}},
  \bibinfo{author}{\bibfnamefont{M.}~\bibnamefont{{Kramer}}},
  \bibinfo{author}{\bibfnamefont{A.}~\bibnamefont{{Possenti}}},
  \bibinfo{author}{\bibfnamefont{R.~N.} \bibnamefont{{Manchester}}},
  \bibinfo{author}{\bibfnamefont{F.}~\bibnamefont{{Camilo}}},
  \bibinfo{author}{\bibfnamefont{M.~A.} \bibnamefont{{McLaughlin}}},
  \bibinfo{author}{\bibfnamefont{D.~R.} \bibnamefont{{Lorimer}}},
  \bibinfo{author}{\bibfnamefont{N.}~\bibnamefont{{D'Amico}}},
  \bibinfo{author}{\bibfnamefont{B.~C.} \bibnamefont{{Joshi}}},
  \bibnamefont{et~al.}, \bibinfo{journal}{Science}
  \textbf{\bibinfo{volume}{303}}, \bibinfo{pages}{1153} (\bibinfo{year}{2004}),
  \eprint{astro-ph/0401086}.

\bibitem[{\citenamefont{{Valtonen} et~al.}(2016)\citenamefont{{Valtonen},
  {Zola}, {Ciprini}, {Gopakumar}, and et~al.}}]{MV2016}
\bibinfo{author}{\bibfnamefont{M.~J.} \bibnamefont{{Valtonen}}},
  \bibinfo{author}{\bibfnamefont{S.}~\bibnamefont{{Zola}}},
  \bibinfo{author}{\bibfnamefont{S.}~\bibnamefont{{Ciprini}}},
  \bibinfo{author}{\bibfnamefont{A.}~\bibnamefont{{Gopakumar}}},
  \bibnamefont{and} \bibinfo{author}{\bibnamefont{et~al.}},
  \bibinfo{journal}{\apjl} \textbf{\bibinfo{volume}{819}}, \bibinfo{eid}{L37}
  (\bibinfo{year}{2016}), \eprint{1603.04171}.

\bibitem[{\citenamefont{Blanchet}(2014)}]{LR_LB}
\bibinfo{author}{\bibfnamefont{L.}~\bibnamefont{Blanchet}},
  \bibinfo{journal}{Living Reviews in Relativity} \textbf{\bibinfo{volume}{17}}
  (\bibinfo{year}{2014}),
  \urlprefix\url{http://www.livingreviews.org/lrr-2014-2}.

\bibitem[{\citenamefont{{Porto} and {Rothstein}}(2017)}]{Porto17}
\bibinfo{author}{\bibfnamefont{R.~A.} \bibnamefont{{Porto}}} \bibnamefont{and}
  \bibinfo{author}{\bibfnamefont{I.~Z.} \bibnamefont{{Rothstein}}},
  \bibinfo{journal}{ArXiv e-prints}  (\bibinfo{year}{2017}),
  \eprint{1703.06433}.

\bibitem[{\citenamefont{{Damour} and {Jaranowski}}(2017)}]{DJ17}
\bibinfo{author}{\bibfnamefont{T.}~\bibnamefont{{Damour}}} \bibnamefont{and}
  \bibinfo{author}{\bibfnamefont{P.}~\bibnamefont{{Jaranowski}}},
  \bibinfo{journal}{ArXiv e-prints}  (\bibinfo{year}{2017}),
  \eprint{1701.02645}.

\bibitem[{\citenamefont{{Foffa} et~al.}(2016)\citenamefont{{Foffa},
  {Mastrolia}, {Sturani}, and {Sturm}}}]{Foffa16}
\bibinfo{author}{\bibfnamefont{S.}~\bibnamefont{{Foffa}}},
  \bibinfo{author}{\bibfnamefont{P.}~\bibnamefont{{Mastrolia}}},
  \bibinfo{author}{\bibfnamefont{R.}~\bibnamefont{{Sturani}}},
  \bibnamefont{and} \bibinfo{author}{\bibfnamefont{C.}~\bibnamefont{{Sturm}}},
  \bibinfo{journal}{ArXiv e-prints}  (\bibinfo{year}{2016}),
  \eprint{1612.00482}.

\bibitem[{\citenamefont{{Bernard} et~al.}(2017)\citenamefont{{Bernard},
  {Blanchet}, {Boh{\'e}}, {Faye}, and {Marsat}}}]{BBBFM16}
\bibinfo{author}{\bibfnamefont{L.}~\bibnamefont{{Bernard}}},
  \bibinfo{author}{\bibfnamefont{L.}~\bibnamefont{{Blanchet}}},
  \bibinfo{author}{\bibfnamefont{A.}~\bibnamefont{{Boh{\'e}}}},
  \bibinfo{author}{\bibfnamefont{G.}~\bibnamefont{{Faye}}}, \bibnamefont{and}
  \bibinfo{author}{\bibfnamefont{S.}~\bibnamefont{{Marsat}}},
  \bibinfo{journal}{\prd} \textbf{\bibinfo{volume}{95}}, \bibinfo{eid}{044026}
  (\bibinfo{year}{2017}), \eprint{1610.07934}.

\bibitem[{\citenamefont{{Damour} et~al.}(2016)\citenamefont{{Damour},
  {Jaranowski}, and {Sch{\"a}fer}}}]{DJS2016}
\bibinfo{author}{\bibfnamefont{T.}~\bibnamefont{{Damour}}},
  \bibinfo{author}{\bibfnamefont{P.}~\bibnamefont{{Jaranowski}}},
  \bibnamefont{and}
  \bibinfo{author}{\bibfnamefont{G.}~\bibnamefont{{Sch{\"a}fer}}},
  \bibinfo{journal}{\prd} \textbf{\bibinfo{volume}{93}}, \bibinfo{eid}{084014}
  (\bibinfo{year}{2016}), \eprint{1601.01283}.

\bibitem[{\citenamefont{{Damour} and {Deruelle}}(1985)}]{DD1985}
\bibinfo{author}{\bibfnamefont{T.}~\bibnamefont{{Damour}}} \bibnamefont{and}
  \bibinfo{author}{\bibfnamefont{N.}~\bibnamefont{{Deruelle}}},
  \bibinfo{journal}{Ann.~Inst.~Henri Poincar{\'e} Phys.~Th{\'e}or., Vol.~43,
  No.~1, p.~107 - 132} \textbf{\bibinfo{volume}{43}}, \bibinfo{pages}{107}
  (\bibinfo{year}{1985}).

\bibitem[{\citenamefont{{Damour} and {Schafer}}(1988)}]{DS1988}
\bibinfo{author}{\bibfnamefont{T.}~\bibnamefont{{Damour}}} \bibnamefont{and}
  \bibinfo{author}{\bibfnamefont{G.}~\bibnamefont{{Schafer}}},
  \bibinfo{journal}{Nuovo Cimento B Serie} \textbf{\bibinfo{volume}{101}},
  \bibinfo{pages}{127} (\bibinfo{year}{1988}).

\bibitem[{\citenamefont{{Sch{\"a}fer} and {Wex}}(1993)}]{SW93}
\bibinfo{author}{\bibfnamefont{G.}~\bibnamefont{{Sch{\"a}fer}}}
  \bibnamefont{and} \bibinfo{author}{\bibfnamefont{N.}~\bibnamefont{{Wex}}},
  \bibinfo{journal}{Physics Letters A} \textbf{\bibinfo{volume}{174}},
  \bibinfo{pages}{196} (\bibinfo{year}{1993}).

\bibitem[{\citenamefont{Memmesheimer et~al.}(2004)\citenamefont{Memmesheimer,
  Gopakumar, and Sch{\"{a}}fer}}]{MGS}
\bibinfo{author}{\bibfnamefont{R.-M.} \bibnamefont{Memmesheimer}},
  \bibinfo{author}{\bibfnamefont{A.}~\bibnamefont{Gopakumar}},
  \bibnamefont{and}
  \bibinfo{author}{\bibfnamefont{G.}~\bibnamefont{Sch{\"{a}}fer}},
  \bibinfo{journal}{Phys. Rev. D} \textbf{\bibinfo{volume}{70}},
  \bibinfo{pages}{104011} (\bibinfo{year}{2004}), ISSN
  \bibinfo{issn}{05562821}, \eprint{0407049}.

\bibitem[{\citenamefont{{Hinder} et~al.}(2010)\citenamefont{{Hinder},
  {Herrmann}, {Laguna}, and {Shoemaker}}}]{Hinder_2010}
\bibinfo{author}{\bibfnamefont{I.}~\bibnamefont{{Hinder}}},
  \bibinfo{author}{\bibfnamefont{F.}~\bibnamefont{{Herrmann}}},
  \bibinfo{author}{\bibfnamefont{P.}~\bibnamefont{{Laguna}}}, \bibnamefont{and}
  \bibinfo{author}{\bibfnamefont{D.}~\bibnamefont{{Shoemaker}}},
  \bibinfo{journal}{\prd} \textbf{\bibinfo{volume}{82}}, \bibinfo{eid}{024033}
  (\bibinfo{year}{2010}), \eprint{0806.1037}.

\bibitem[{\citenamefont{{Huerta} et~al.}(2016)\citenamefont{{Huerta}, {Kumar},
  {Agarwal}, {George}, {Schive}, {Pfeiffer}, {Chu}, {Boyle}, {Hemberger},
  {Kidder} et~al.}}]{eIMR}
\bibinfo{author}{\bibfnamefont{E.~A.} \bibnamefont{{Huerta}}},
  \bibinfo{author}{\bibfnamefont{P.}~\bibnamefont{{Kumar}}},
  \bibinfo{author}{\bibfnamefont{B.}~\bibnamefont{{Agarwal}}},
  \bibinfo{author}{\bibfnamefont{D.}~\bibnamefont{{George}}},
  \bibinfo{author}{\bibfnamefont{H.-Y.} \bibnamefont{{Schive}}},
  \bibinfo{author}{\bibfnamefont{H.~P.} \bibnamefont{{Pfeiffer}}},
  \bibinfo{author}{\bibfnamefont{T.}~\bibnamefont{{Chu}}},
  \bibinfo{author}{\bibfnamefont{M.}~\bibnamefont{{Boyle}}},
  \bibinfo{author}{\bibfnamefont{D.~A.} \bibnamefont{{Hemberger}}},
  \bibinfo{author}{\bibfnamefont{L.~E.} \bibnamefont{{Kidder}}},
  \bibnamefont{et~al.}, \bibinfo{journal}{ArXiv e-prints}
  (\bibinfo{year}{2016}), \eprint{1609.05933}.

\bibitem[{\citenamefont{{Damour} et~al.}(2004)\citenamefont{{Damour},
  {Gopakumar}, and {Iyer}}}]{DGI}
\bibinfo{author}{\bibfnamefont{T.}~\bibnamefont{{Damour}}},
  \bibinfo{author}{\bibfnamefont{A.}~\bibnamefont{{Gopakumar}}},
  \bibnamefont{and} \bibinfo{author}{\bibfnamefont{B.~R.}
  \bibnamefont{{Iyer}}}, \bibinfo{journal}{\prd} \textbf{\bibinfo{volume}{70}},
  \bibinfo{eid}{064028} (\bibinfo{year}{2004}), \eprint{gr-qc/0404128}.

\bibitem[{\citenamefont{{K{\"o}nigsd{\"o}rffer} and {Gopakumar}}(2006)}]{KG06}
\bibinfo{author}{\bibfnamefont{C.}~\bibnamefont{{K{\"o}nigsd{\"o}rffer}}}
  \bibnamefont{and}
  \bibinfo{author}{\bibfnamefont{A.}~\bibnamefont{{Gopakumar}}},
  \bibinfo{journal}{\prd} \textbf{\bibinfo{volume}{73}}, \bibinfo{eid}{124012}
  (\bibinfo{year}{2006}), \eprint{gr-qc/0603056}.

\bibitem[{\citenamefont{{Tanay} et~al.}(2016)\citenamefont{{Tanay}, {Haney},
  and {Gopakumar}}}]{THG16}
\bibinfo{author}{\bibfnamefont{S.}~\bibnamefont{{Tanay}}},
  \bibinfo{author}{\bibfnamefont{M.}~\bibnamefont{{Haney}}}, \bibnamefont{and}
  \bibinfo{author}{\bibfnamefont{A.}~\bibnamefont{{Gopakumar}}},
  \bibinfo{journal}{\prd} \textbf{\bibinfo{volume}{93}}, \bibinfo{eid}{064031}
  (\bibinfo{year}{2016}), \eprint{1602.03081}.

\bibitem[{\citenamefont{{Damour} and {Deruelle}}(1986)}]{DD86}
\bibinfo{author}{\bibfnamefont{T.}~\bibnamefont{{Damour}}} \bibnamefont{and}
  \bibinfo{author}{\bibfnamefont{N.}~\bibnamefont{{Deruelle}}},
  \bibinfo{journal}{Ann.~Inst.~Henri Poincar{\'e} Phys.~Th{\'e}or., Vol.~44,
  No.~3, p.~263 - 292} \textbf{\bibinfo{volume}{44}}, \bibinfo{pages}{263}
  (\bibinfo{year}{1986}).

\bibitem[{\citenamefont{{Damour} and {Taylor}}(1992)}]{DT92}
\bibinfo{author}{\bibfnamefont{T.}~\bibnamefont{{Damour}}} \bibnamefont{and}
  \bibinfo{author}{\bibfnamefont{J.~H.} \bibnamefont{{Taylor}}},
  \bibinfo{journal}{\prd} \textbf{\bibinfo{volume}{45}}, \bibinfo{pages}{1840}
  (\bibinfo{year}{1992}).

\bibitem[{\citenamefont{{Stairs}}(2003)}]{IS03_LR}
\bibinfo{author}{\bibfnamefont{I.~H.} \bibnamefont{{Stairs}}},
  \bibinfo{journal}{Living Reviews in Relativity} \textbf{\bibinfo{volume}{6}},
  \bibinfo{eid}{5} (\bibinfo{year}{2003}), \eprint{astro-ph/0307536}.

\bibitem[{\citenamefont{{De Vittori} et~al.}(2014)\citenamefont{{De Vittori},
  {Gopakumar}, {Gupta}, and {Jetzer}}}]{LGGJ}
\bibinfo{author}{\bibfnamefont{L.}~\bibnamefont{{De Vittori}}},
  \bibinfo{author}{\bibfnamefont{A.}~\bibnamefont{{Gopakumar}}},
  \bibinfo{author}{\bibfnamefont{A.}~\bibnamefont{{Gupta}}}, \bibnamefont{and}
  \bibinfo{author}{\bibfnamefont{P.}~\bibnamefont{{Jetzer}}},
  \bibinfo{journal}{\prd} \textbf{\bibinfo{volume}{90}}, \bibinfo{eid}{124066}
  (\bibinfo{year}{2014}), \eprint{1410.6311}.

\bibitem[{\citenamefont{{Kocsis} et~al.}(2006)\citenamefont{{Kocsis},
  {G{\'a}sp{\'a}r}, and {M{\'a}rka}}}]{KGM06}
\bibinfo{author}{\bibfnamefont{B.}~\bibnamefont{{Kocsis}}},
  \bibinfo{author}{\bibfnamefont{M.~E.} \bibnamefont{{G{\'a}sp{\'a}r}}},
  \bibnamefont{and}
  \bibinfo{author}{\bibfnamefont{S.}~\bibnamefont{{M{\'a}rka}}},
  \bibinfo{journal}{\apj} \textbf{\bibinfo{volume}{648}}, \bibinfo{pages}{411}
  (\bibinfo{year}{2006}), \eprint{astro-ph/0603441}.

\bibitem[{\citenamefont{{Garcia-Bellido} and {Nesseris}}(2017)}]{BN17}
\bibinfo{author}{\bibfnamefont{J.}~\bibnamefont{{Garcia-Bellido}}}
  \bibnamefont{and}
  \bibinfo{author}{\bibfnamefont{S.}~\bibnamefont{{Nesseris}}},
  \bibinfo{journal}{ArXiv e-prints}  (\bibinfo{year}{2017}),
  \eprint{1711.09702}.

\bibitem[{\citenamefont{{Hansen}}(1972)}]{H72}
\bibinfo{author}{\bibfnamefont{R.~O.} \bibnamefont{{Hansen}}},
  \bibinfo{journal}{\prd} \textbf{\bibinfo{volume}{5}}, \bibinfo{pages}{1021}
  (\bibinfo{year}{1972}).

\bibitem[{\citenamefont{{Walker} and {Will}}(1979)}]{WW79}
\bibinfo{author}{\bibfnamefont{M.}~\bibnamefont{{Walker}}} \bibnamefont{and}
  \bibinfo{author}{\bibfnamefont{C.~M.} \bibnamefont{{Will}}},
  \bibinfo{journal}{\prd} \textbf{\bibinfo{volume}{19}}, \bibinfo{pages}{3483}
  (\bibinfo{year}{1979}).

\bibitem[{\citenamefont{{O'Leary} et~al.}(2009)\citenamefont{{O'Leary},
  {Kocsis}, and {Loeb}}}]{OKL09}
\bibinfo{author}{\bibfnamefont{R.~M.} \bibnamefont{{O'Leary}}},
  \bibinfo{author}{\bibfnamefont{B.}~\bibnamefont{{Kocsis}}}, \bibnamefont{and}
  \bibinfo{author}{\bibfnamefont{A.}~\bibnamefont{{Loeb}}},
  \bibinfo{journal}{\mnras} \textbf{\bibinfo{volume}{395}},
  \bibinfo{pages}{2127} (\bibinfo{year}{2009}), \eprint{0807.2638}.

\bibitem[{\citenamefont{{Tsang}}(2013)}]{DT13}
\bibinfo{author}{\bibfnamefont{D.}~\bibnamefont{{Tsang}}},
  \bibinfo{journal}{\apj} \textbf{\bibinfo{volume}{777}}, \bibinfo{eid}{103}
  (\bibinfo{year}{2013}), \eprint{1307.3554}.

\bibitem[{\citenamefont{{Mikkola}}(1987)}]{Mikkola87}
\bibinfo{author}{\bibfnamefont{S.}~\bibnamefont{{Mikkola}}},
  \bibinfo{journal}{Celestial Mechanics} \textbf{\bibinfo{volume}{40}},
  \bibinfo{pages}{329} (\bibinfo{year}{1987}).

\bibitem[{\citenamefont{{Blanchet} and {Schaefer}}(1989)}]{BS89}
\bibinfo{author}{\bibfnamefont{L.}~\bibnamefont{{Blanchet}}} \bibnamefont{and}
  \bibinfo{author}{\bibfnamefont{G.}~\bibnamefont{{Schaefer}}},
  \bibinfo{journal}{\mnras} \textbf{\bibinfo{volume}{239}},
  \bibinfo{pages}{845} (\bibinfo{year}{1989}).

\bibitem[{\citenamefont{{Junker} and {Schaefer}}(1992)}]{JS92}
\bibinfo{author}{\bibfnamefont{W.}~\bibnamefont{{Junker}}} \bibnamefont{and}
  \bibinfo{author}{\bibfnamefont{G.}~\bibnamefont{{Schaefer}}},
  \bibinfo{journal}{\mnras} \textbf{\bibinfo{volume}{254}},
  \bibinfo{pages}{146} (\bibinfo{year}{1992}).

\bibitem[{\citenamefont{{Tiwari} et~al.}(2016)\citenamefont{{Tiwari},
  {Klimenko}, {Christensen}, {Huerta}, {Mohapatra}, {Gopakumar}, {Haney},
  {Ajith}, {McWilliams}, {Vedovato} et~al.}}]{Tiwari16}
\bibinfo{author}{\bibfnamefont{V.}~\bibnamefont{{Tiwari}}},
  \bibinfo{author}{\bibfnamefont{S.}~\bibnamefont{{Klimenko}}},
  \bibinfo{author}{\bibfnamefont{N.}~\bibnamefont{{Christensen}}},
  \bibinfo{author}{\bibfnamefont{E.~A.} \bibnamefont{{Huerta}}},
  \bibinfo{author}{\bibfnamefont{S.~R.~P.} \bibnamefont{{Mohapatra}}},
  \bibinfo{author}{\bibfnamefont{A.}~\bibnamefont{{Gopakumar}}},
  \bibinfo{author}{\bibfnamefont{M.}~\bibnamefont{{Haney}}},
  \bibinfo{author}{\bibfnamefont{P.}~\bibnamefont{{Ajith}}},
  \bibinfo{author}{\bibfnamefont{S.~T.} \bibnamefont{{McWilliams}}},
  \bibinfo{author}{\bibfnamefont{G.}~\bibnamefont{{Vedovato}}},
  \bibnamefont{et~al.}, \bibinfo{journal}{\prd} \textbf{\bibinfo{volume}{93}},
  \bibinfo{eid}{043007} (\bibinfo{year}{2016}), \eprint{1511.09240}.

\bibitem[{\citenamefont{{Boetzel} et~al.}(2018)\citenamefont{{Boetzel}, {Cho},
  and {Gopakumar}}}]{BCG18}
\bibinfo{author}{\bibfnamefont{Y.}~\bibnamefont{{Boetzel}}},
  \bibinfo{author}{\bibfnamefont{G.}~\bibnamefont{{Cho}}}, \bibnamefont{and}
  \bibinfo{author}{\bibfnamefont{A.}~\bibnamefont{{Gopakumar}}},
  \bibinfo{journal}{To be published}  (\bibinfo{year}{2018}).

\bibitem[{\citenamefont{{Hopper} and {Cardoso}}(2017)}]{HC17}
\bibinfo{author}{\bibfnamefont{S.}~\bibnamefont{{Hopper}}} \bibnamefont{and}
  \bibinfo{author}{\bibfnamefont{V.}~\bibnamefont{{Cardoso}}},
  \bibinfo{journal}{ArXiv e-prints}  (\bibinfo{year}{2017}),
  \eprint{1706.02791}.

\bibitem[{\citenamefont{{Bini} and {Geralico}}(2017)}]{BG17}
\bibinfo{author}{\bibfnamefont{D.}~\bibnamefont{{Bini}}} \bibnamefont{and}
  \bibinfo{author}{\bibfnamefont{A.}~\bibnamefont{{Geralico}}},
  \bibinfo{journal}{General Relativity and Gravitation}
  \textbf{\bibinfo{volume}{49}}, \bibinfo{eid}{84} (\bibinfo{year}{2017}),
  \eprint{1707.09814}.

\bibitem[{\citenamefont{{Hopper}}(2017)}]{H17}
\bibinfo{author}{\bibfnamefont{S.}~\bibnamefont{{Hopper}}},
  \bibinfo{journal}{ArXiv e-prints}  (\bibinfo{year}{2017}),
  \eprint{1706.05455}.

\bibitem[{\citenamefont{{Bini} and {Damour}}(2012)}]{BD_2012}
\bibinfo{author}{\bibfnamefont{D.}~\bibnamefont{{Bini}}} \bibnamefont{and}
  \bibinfo{author}{\bibfnamefont{T.}~\bibnamefont{{Damour}}},
  \bibinfo{journal}{\prd} \textbf{\bibinfo{volume}{86}}, \bibinfo{eid}{124012}
  (\bibinfo{year}{2012}), \eprint{1210.2834}.

\bibitem[{\citenamefont{Damour}(2016)}]{Damour16}
\bibinfo{author}{\bibfnamefont{T.}~\bibnamefont{Damour}},
  \bibinfo{journal}{Phys. Rev. D} \textbf{\bibinfo{volume}{94}},
  \bibinfo{pages}{104015} (\bibinfo{year}{2016}),
  \urlprefix\url{https://link.aps.org/doi/10.1103/PhysRevD.94.104015}.

\bibitem[{\citenamefont{{Bini} and {Damour}}(2017)}]{BD17}
\bibinfo{author}{\bibfnamefont{D.}~\bibnamefont{{Bini}}} \bibnamefont{and}
  \bibinfo{author}{\bibfnamefont{T.}~\bibnamefont{{Damour}}},
  \bibinfo{journal}{\prd} \textbf{\bibinfo{volume}{96}}, \bibinfo{eid}{064021}
  (\bibinfo{year}{2017}), \eprint{1706.06877}.

\end{thebibliography}
\bibliographystyle{apsrev}

\end{document}